\documentclass[11pt]{article}
\pdfoutput=1
\usepackage{jcapmod}

\usepackage{booktabs}
\usepackage[english]{babel}
\usepackage{amsmath, amsfonts, amssymb, amsbsy, amstext, amsthm}
\usepackage{hyperref}
\usepackage{graphicx}
\usepackage{siunitx}
\usepackage[caption=false]{subfig}

\allowdisplaybreaks[1]

\numberwithin{equation}{section}

\def\beq{\begin{equation}}
\def\eeq{\end{equation}}

\def\d{{\rm d}}
\def\i{{\rm i}}
\def\Neff{N_{\rm eff}}
\def\Tr{T_R}
\def\Tf{T_F}
\def\Tff{T_{\tilde F}}
\def\Tfl{T_{\rm fluid}}
\def\Trec{T_{\rm rec}}
\def\Teq{T_{\rm eq}}
\def\Td{T_D}
\def\mt{m_t}
\def\Mp{M_{\rm pl}}
\def\L{\mathcal{L}}
\def\M{\mathcal{M}}
\def\O{\mathcal{O}}

\DeclareRobustCommand{\SkipTocEntry}[4]{}

\begin{document}

\pagenumbering{roman}
\begin{titlepage}
\baselineskip=15.5pt \thispagestyle{empty}

\bigskip\

\vspace{1cm}
\begin{center}
{\fontsize{20.74}{24}\selectfont \sffamily \bfseries A New Target for Cosmic Axion Searches}
\end{center}

\vspace{0.2cm}
\begin{center}
{\fontsize{12}{30}\selectfont Daniel Baumann,$^{\bigstar,\spadesuit}$ Daniel Green$^{\blacklozenge,\clubsuit}$ and Benjamin Wallisch$^{\bigstar}$} 
\end{center}

\begin{center}
\vskip8pt
\textsl{$^{\bigstar}$ Department of Applied Mathematics and Theoretical Physics,\\University of Cambridge, Cambridge, CB3 0WA, UK}

\vskip8pt
\textsl{$^\spadesuit$ Institute of Theoretical Physics, University of Amsterdam,\\Science Park, Amsterdam, 1090 GL, The Netherlands}

\vskip7pt
\textsl{$^\blacklozenge$ Department of Physics, University of California, Berkeley, CA 94720, US}

\vskip7pt
\textsl{$^\clubsuit$ Canadian Institute for Theoretical Astrophysics, Toronto, ON M5S 3H8, Canada}
\end{center}

\vspace{1.2cm}
\hrule \vspace{0.3cm}
\noindent {\sffamily \bfseries Abstract}\\[0.1cm]
Future cosmic microwave background experiments have the potential to probe the density of relativistic species at the subpercent level. This sensitivity allows light thermal relics to be detected up to arbitrarily high decoupling temperatures. Conversely, the absence of a detection would require extra light species never to have been in equilibrium with the Standard Model. In this paper, we exploit this feature to demonstrate the sensitivity of future cosmological observations to the couplings of axions to all of the Standard Model degrees of freedom.  In many cases, the constraints achievable from cosmology will surpass existing bounds from laboratory experiments and astrophysical observations by orders of magnitude.
\vskip10pt
\hrule
\vskip10pt

\vspace{0.6cm}
\end{titlepage}

\thispagestyle{empty}
\setcounter{page}{2}
\tableofcontents

\clearpage
\pagenumbering{arabic}
\setcounter{page}{1}

%%%%%%%%%%%%%%%%
\section{Introduction}
\label{sec:intro}
%%%%%%%%%%%%%%%%

Most of what we know about the history of the universe comes from the observations of light emitted at or after recombination.  To learn about earlier times we rely either on theoretical extrapolations or the observations of relics that are left over from an earlier period. One of the most remarkable results of the Planck satellite is the detection of free-streaming cosmic neutrinos~\cite{Ade:2015xua, Follin:2015hya, Baumann:2015rya}, with an energy density that is consistent with the predicted freeze-out abundance created one second after the Big Bang. Probing even earlier times requires detecting new particles that are more weakly coupled than neutrinos.  Such particles arise naturally in many extensions of the Standard Model~(SM)~\cite{Jaeckel:2010ni, Essig:2013lka}.  Particularly well-motivated are Goldstone bosons created by the spontaneous breaking of additional global symmetries. The scale of the symmetry breaking then determines the strength of the coupling to the SM.  If this scale is sufficiently high, then these particles can escape detection at colliders, but cosmology will still be sensitive to them. 

\vskip6pt
Goldstone bosons are either massless (if the broken symmetry was exact) or naturally light (if it was approximate). Examples of light pseudo-Nambu-Goldstone bosons (pNGBs) are {\it axions}~\cite{1977PhRvL..38.1440P, Weinberg:1977ma, 1978PhRvL..40..279W}, {\it familons}~\cite{PhysRevLett.49.1549, reiss1982can, kim1987light}, and {\it majorons}~\cite{Chikashige:1980ui, Chikashige:1980qk}, associated with spontaneously broken Peccei-Quinn, family and lepton-number symmetry, respectively.  Below the scale of the spontaneous symmetry breaking (SSB), the couplings of  the Goldstone bosons $\phi$ to the SM degrees of freedom can be characterized through a set of effective interactions
\beq
\frac{\O_\phi\hskip1pt \O_{\rm SM}}{\Lambda^\Delta}\, , \label{equ:coupling}
\eeq
where $\Lambda$ is related to the symmetry breaking scale.  Axion, familon and majoron models are characterized by different couplings in~(\ref{equ:coupling}).\footnote{We will follow the common practice of reserving the name axion or axion-like particle (ALP) for pNGBs that couple to the gauge bosons of the SM through operators like $\phi F_{\mu \nu}\tilde F^{\mu \nu}$.}  These couplings are constrained by laboratory experiments~\cite{Essig:2013lka,Graham:2015ouw}, by astrophysics~\cite{Raffelt:1996wa, Raffelt:2012kt} and by cosmology~\cite{Brust:2013xpv, Marsh:2015xka}.  While laboratory constraints have the advantage of being direct measurements, their main drawback is that they are usually rather model-specific and sensitive only to narrow windows of pNGB masses.  Astrophysical and cosmological constraints are complimentary since they are relatively insensitive to the detailed form of the couplings to the SM and span a wide range of masses.  The main astrophysical constraints on new light particles come from stellar cooling~\cite{Raffelt:1996wa}. In order not to disrupt successful models of stellar evolution, any new light particles must be more weakly coupled than neutrinos.  Moreover, since neutrinos couple to the rest of the SM through a dimension-six operator (suppressed by the electroweak scale), the constraints on extra particles are particularly severe for dimension-four and dimension-five couplings to the SM.

\vskip8pt
In this paper, we will show that cosmology is remarkably sensitive to extra light particles.  This is because interactions like~(\ref{equ:coupling}) can bring these particles into equilibrium  with the SM particles. Moreover, thermal equilibrium is democratic.  Any new light field that was in thermal equilibrium in the past will have a number density that is comparable to that of photons.  This is the reason why neutrinos have been detected with high significance in the CMB~\cite{Ade:2015xua, Follin:2015hya, Baumann:2015rya} despite their weak coupling. Like astrophysical constraints, cosmology therefore requires any new light particles to be more weakly coupled than neutrinos.   Given the Moore's law-like improvements in CMB detector sensitivity~\cite{Abazajian:2013oma, Wu:2014hta}, cosmology will push the sensitivity to new light particles beyond the strength of weak scale interactions and has the potential to explore a fundamentally new territory of physics beyond the SM.
\begin{figure}[t!]
\begin{center}
\includegraphics[width=0.7\textwidth]{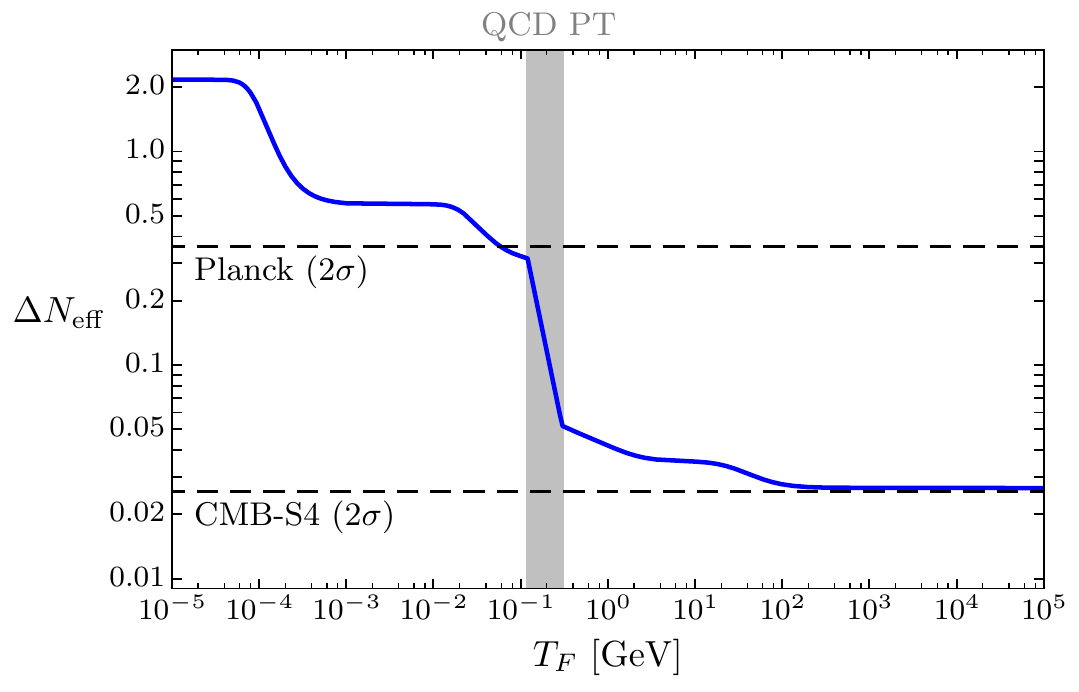}
\caption{Contribution of a single thermally-decoupled Goldstone boson to the effective number of neutrinos, $\Delta \Neff$, as a function of the freeze-out temperature $\Tf$. Shown are also the current $2\sigma$ sensitivity of the Planck satellite~\cite{Ade:2015xua} and an (optimistic) estimate of the sensitivity of a future CMB-S4 mission~\cite{Baumann:2015rya}.  }
\label{fig:freezeout}
\end{center}
\end{figure}

\vskip6pt
The total energy density in relativistic species is often defined as 
\beq
\rho_r = \left[ 1+ \frac{7}{8} \left(\frac{4}{11}\right)^{\!4/3} \Neff\right] \rho_\gamma\, , \label{equ:Neff}
\eeq
where $\rho_\gamma$ is the energy density of photons and the parameter $\Neff$ is often called the effective number of neutrinos,  
although there may be contributions to $\Neff$ that have nothing to do with neutrinos~(see e.g.~\cite{Weinberg:2013kea}). The SM predicts $\Neff = 3.046$ from neutrinos~\cite{Mangano:2005cc} and the current constraint from the Planck satellite is $\Neff =  3.04 \pm 0.18$~\cite{Ade:2015xua}. Figure~\ref{fig:freezeout} shows the extra contribution to the radiation density of a thermally-decoupled Goldstone boson as a function of its freeze-out temperature~$\Tf$. We see that particles that decoupled {\it after} the QCD phase transition are ruled out (or at least are highly constrained) by the observations of the Planck satellite~\cite{Brust:2013xpv}. On the other hand, the effect of particles that decoupled {\it before} the QCD phase transition is suppressed by an order of magnitude, $0.05 \ge \Delta \Neff \ge 0.027$. Although Planck is blind to these particles, this regime is within reach of future experiments.  In particular, the planned CMB Stage IV (CMB-S4) experiments have the potential to constrain (or detect) extra relativistic species at the level of $\sigma(\Neff) \sim 0.01$~\cite{Abazajian:2013oma, Wu:2014hta, Baumann:2015rya}.

\vskip6pt
The fact that the minimal thermal contribution may be detectable has interesting consequences.  First, the level $\Delta \Neff = 0.027$ provides a natural observational target (see e.g.~\cite{Brust:2013xpv, Salvio:2013iaa, Kawasaki:2015ofa,Chacko:2015noa} for related discussions). Second, even the absence of a detection would be very informative, because it would strongly constrain the couplings between the extra light relics and the SM degrees of freedom. This is because a thermal abundance can be avoided\hskip1pt\footnote{A thermal abundance may be diluted below the level of Fig.~\ref{fig:freezeout} if extra massive particles are added to the SM.  However, a significant change to our conclusions would require a very large number of new particles or a significant amount of non-equilibrium photon production. In addition, the possibility that dark sectors never reach thermal equilibrium with the SM (see e.g.~\cite{Berezhiani:1995am, Hodges:1993yb, Feng:2008mu, Foot:2014uba, Reece:2015lch}) is strongly constrained by the physics of reheating~\cite{Adshead:2016xxj}. } if the reheating temperature of the universe, $\Tr$, is below the would-be freeze-out temperature, i.e.~$\Tr < \Tf$. In that case, the extra particles have never been in thermal equilibrium and their densities therefore do not have to be detectable.  In the absence of a detection, requiring $\Tf(\Lambda) > \Tr$ would place very strong bounds on the scale(s) in~(\ref{equ:coupling}), i.e.~$\Lambda > \Tf^{-1}(\Tr)$. As we will see, in many cases the cosmological bounds will be much stronger than existing bounds from laboratory experiments and astrophysical observations. We note that these constraints make no assumption about the nature of dark matter because the thermal population of axions arises independently of a possible cold population. On the other hand, we have to assume that the effective description of the pNGBs with interactions of the form of~(\ref{equ:coupling}) holds up to $\Tf\ll\Lambda$. This is equivalent to assuming that the UV completion of the effective theory is not too weakly coupled. Moreover, we also require the absence of any significant dilution of $\Delta\Neff$ after freeze-out. In practice, this means that we are restricting to scenarios in which the number of additional relativistic degrees of freedom at the freeze-out temperature is bounded by  $\Delta{g_*(\Tf)}\lesssim{g_{*}^{\rm SM}(\Tf)}\approx\num{e2}$.

\vskip4pt
The couplings of pNGBs to SM fermions $\psi$ can lead to a more complicated thermal evolution than the simple freeze-out scenario. Below the scale of electroweak symmetry breaking (EWSB), the approximate chiral symmetry of the fermions makes the interactions with the pNGBs effectively marginal. The temperature dependence of the interaction rate is then weaker than that of the Hubble expansion rate, leading to a recoupling (i.e.~freeze-in) of the pNGBs at low temperatures.  To avoid a large density of pNGBs requires that the freeze-in temperature $\Tff$ is smaller than the mass of the fermions participating in the interactions, $\Tff < m_\psi$, so that the interaction rate becomes Boltzmann suppressed before freeze-in can occur. Again, this constraint can be expressed as a bound on the scale(s) $\Lambda$ that couple the pNGBs to the SM fermions. Although the freeze-in constraints are usually weaker than the freeze-out constraints, they have the advantage that they do not make any assumptions about the reheating temperature (as long as reheating occurs above~$T \sim m_\psi$). Moreover, freeze-in produces larger contributions to $\Delta \Neff$ which are  detectable with a less sensitive experiment. 

\vskip10pt
In the rest of this paper, we will show that cosmology is highly sensitive to axions, and other pNGBs, when $\Delta \Neff = 0.027$ is detectable.  To simplify the narrative, we will assume that this sensitivity will be reached with CMB-S4, either on its own or in conjunction with other data~\cite{Font-Ribera:2013rwa,Manzotti:2015ozr}.  Alternatively, our arguments could be viewed as strong motivation for reaching this critical level of sensitivity in future experiments.  In the following, we will derive  bounds on the couplings of pNGBs to the SM arising from the absence of a detection.  We will assume the mass range $0 \leq m_\phi < \SI{1}{MeV}$, so that the only possible decays of the pNGBs are to photons or neutrinos.   This regime is probed by measurements of $\Neff$ for $m_\phi \leq \Trec$ and by warm dark matter constraints for $m_\phi > \Trec$ (see e.g.~\cite{Archidiacono:2015mda, DiValentino:2015wba}), where $\Trec  \approx \SI{0.26}{eV}$ is the temperature at recombination. We will discuss in turn the couplings to gauge bosons~(\S\ref{sec:axions}), to fermions~(\S\ref{sec:fermions}) and to neutrinos~(\S\ref{sec:neutrinos}). The corresponding interaction rates are computed in Appendix~\ref{app:rates} and the effects of decays are discussed in Appendix~\ref{app:decays}.

%%%%%%%%%%%%%%%%%%%
\section{Constraints on Axions}
\label{sec:axions}
%%%%%%%%%%%%%%%%%%%

Axions arise naturally in many areas of high-energy physics, the QCD axion being a particularly well-motivated example.  Besides providing a solution to the strong CP problem~\cite{1977PhRvL..38.1440P, Weinberg:1977ma, 1978PhRvL..40..279W}, the QCD axion also serves as a natural dark matter candidate~\cite{Preskill:1982cy, Abbott:1982af, Dine:1982ah}.  Moreover, light axions appear prolifically in string theory~\cite{svrcek2006axions, Arvanitaki:2009fg, Baumann:2014nda} and have been proposed to explain the small mass of the inflaton~\cite{Freese:1990rb} as well as to solve the hierarchy problem~\cite{Graham:2015cka}.  Finally, axions are a compelling example of a new particle that is experimentally elusive~\cite{Graham:2015ouw, Essig:2013lka} because of its weak coupling rather than due to kinematic constraints. 
  
\vskip4pt
What typically distinguishes axions from other pNGBs are their unique couplings to the SM gauge fields.  Prior to EWSB, we consider the following effective theory with shift-symmetric couplings of the axion to the SM gauge sector:
\beq
\L_{\phi{\rm EW}} = - \frac{1}{4}\frac{\phi}{\Lambda} \left( c_1 \hskip1pt B_{\mu\nu} \tilde{B}^{\mu\nu} + c_2 \hskip1pt W_{\mu\nu}^a \tilde{W}^{\mu\nu,a} +c_3 \hskip1pt G_{\mu\nu}^a \tilde{G}^{\mu\nu,a} \right)  ,  \label{equ:LEW0}
\eeq
where $X_{\mu\nu} \equiv  \{B_{\mu\nu}, W_{\mu\nu}^a, G_{\mu\nu}^a\}$ are the field strengths associated with the gauge groups $\{U(1)_Y,$ $SU(2)_L, SU(3)_c\}$, and $\tilde{X}^{\mu\nu} \equiv \frac{1}{2} \epsilon^{\mu\nu\rho\sigma} X_{\rho\sigma}$ are their duals. Axion models will typically include couplings to all SM gauge fields, but only the coupling to gluons is strictly necessary to solve the strong CP problem.

\vskip4pt
At high energies, the rate of axion production through the gauge field interactions in~(\ref{equ:LEW0}) can be expressed as~\cite{Salvio:2013iaa} (see also~\cite{Braaten:1991dd, Bolz:2000fu, Masso:2002np, Graf:2010tv})
\beq
\Gamma(T,\Lambda_n) = \sum_{n=1}^3\gamma_n(T)\, \frac{T^3}{\Lambda_n^2}\, , \label{eqn:Gamma}
\eeq
where $\Lambda_n \equiv \Lambda/c_n$.  The prefactors $\gamma_n(T)$ have their origin in the running of the couplings and are only weakly dependent on temperature. For simplicity of presentation, we will treat these functions as constants in the main text, but take them into account in Appendix~\ref{app:rates}. We see that the production rate, $\Gamma \propto T^3$, decreases faster than the expansion rate during the radiation era, $H \propto T^2$. The axions therefore {\it freeze out} when the production rate drops below the expansion rate, with the freeze-out temperature $\Tf$ determined by $\Gamma(\Tf) = H(\Tf)$.  This thermal abundance can be avoided if the reheating temperature of the universe $\Tr$ was below the would-be freeze-out temperature, i.e.~$\Tr < \Tf$. In that case, the temperature of the universe was simply never high enough to bring the axions into thermal equilibrium.  We can express this condition as
\beq
\Gamma(\Tr,\Lambda_n) < H(\Tr) = \frac{\pi}{\sqrt{90}} \sqrt{g_{*,R}}\,\frac{\Tr^2}{\Mp}\, ,
\eeq
where $\Mp$ is the reduced Planck mass and $g_{*,R} \equiv g_*(\Tr)$ denotes the effective number of relativistic species at $\Tr$. For a given reheating temperature, this is a constraint on the couplings~$\Lambda_n$ in~(\ref{eqn:Gamma}). Treating the different axion couplings separately, we can write
\beq
\Lambda_n \, >\, \left(\frac{\pi^2}{90} g_{*,R}\right)^{\!-1/4} \sqrt{\gamma_{n,R} \hskip1pt \Tr\hskip1pt \Mp}\, ,
\label{eq:freezeoutconstraint}
\eeq
where $\gamma_{n,R} \equiv \gamma_n(\Tr)$. In the following, we will evaluate these bounds for the couplings to photons~(\S\ref{ssec:photons}) and gluons~(\S\ref{ssec:gluons}), and compare them to existing laboratory and astrophysical constraints.

%%%%%%%%%%%%%%%%%%%%%
\subsection{Coupling to Photons}
\label{ssec:photons}
%%%%%%%%%%%%%%%%%%%%%

\begin{figure}[t!]
\begin{center}
\includegraphics[width=0.7\textwidth]{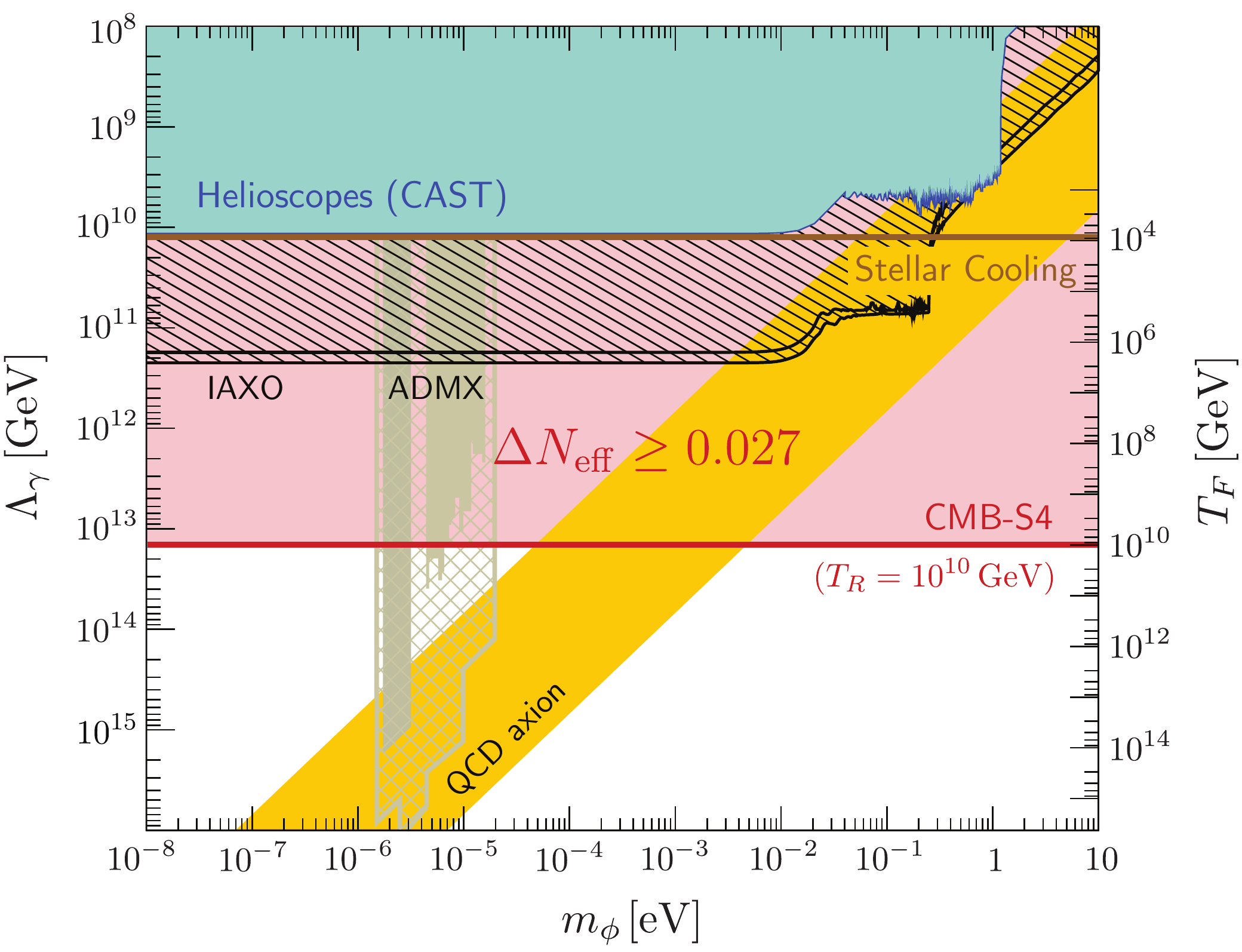}
\caption{Comparison between current constraints on the axion-photon coupling and the sensitivity of a future CMB-S4 mission (figure adapted from~\cite{Carosi:2013rla}).  Future laboratory constraints (IAXO and ADMX) are shown as shaded regions. The yellow band indicates a range of representative models for the QCD axion (not assuming that it provides all of the dark matter). The future CMB bound is a function of the reheating temperature $\Tr$ and the displayed constraint conservatively assumes that the photon coupling derives only from the coupling to $U(1)_Y$ above the electroweak scale. Specific axion models typically also involve a coupling to $SU(2)_L$ in which case the bound would strengthen by an order of magnitude or more (see Appendix~\ref{app:rates}).  We note that ADMX assumes that the axion is all of the dark matter, while all other constraints do not have this restriction.}
\label{fig:S4axion}
\end{center}
\end{figure}

The operator that has been most actively investigated experimentally is the coupling to photons,
\beq
\hspace{-1.9cm}\L_{\phi{\rm EW}} \supset \L_{\phi \gamma} = - \frac{1}{4}\frac{\phi}{\Lambda_\gamma}  F_{\mu \nu} \tilde F^{\mu \nu}\, , \label{equ:phiF}
\eeq
where $F^{\mu\nu}$ is the electromagnetic field strength and $\tilde{F}^{\mu\nu}$ is its dual. The electroweak couplings $\Lambda_1$ and $\Lambda_2$ are related to the photon coupling $\Lambda_\gamma$ via $\Lambda_\gamma^{-1} = \cos^2\theta_w \hskip1pt\Lambda_1^{-1} + \sin^2\theta_w \hskip1pt\Lambda_2^{-1}$, where $\theta_w \approx 30^\circ$ is the Weinberg mixing angle.  Photons are easily produced in large numbers in both the laboratory and in many astrophysical settings which makes this coupling a particularly fruitful target for axion searches.  

\vskip4pt
In Appendix~\ref{app:rates}, we show in detail how the constraints~(\ref{eq:freezeoutconstraint}) on the couplings to the electroweak gauge bosons map into a constraint on the coupling to photons. This constraint is a function of the relative size of the couplings to the $SU(2)_L$ and $U(1)_Y$ sectors, as measured by the ratio $c_2/c_1$ in~(\ref{equ:LEW0}). To be conservative, we will here present the weakest constraint which arises for $c_2=0$ when the axion only couples to the $U(1)_Y$ gauge field. A specific axion model is likely to also couple to the $SU(2)_L$ sector, i.e.~have $c_2\ne 0$, and the constraint on $\Lambda_\gamma$ would then be stronger (as can be seen explicitly in Appendix~\ref{app:rates}). Using $\gamma_{1,R} \approx \gamma_1(\SI{e10}{GeV}) = 0.017$ and $g_{*,R} = 106.75+1$, we find
\begin{align}
\Lambda_\gamma &\,>\, \SI{1.4e13}{GeV} \left(\frac{\Tr}{\SI{e10}{GeV}}\right)^{\!1/2}\, . \label{equ:PhotonBound}
\end{align}
For a reheating temperature of about \SI{e10}{GeV}, the bound in~(\ref{equ:PhotonBound}) is three orders of magnitude stronger than the best current constraints (cf.~Fig.~\ref{fig:S4axion}).  Even for a reheating temperature as low as $\SI{e4}{GeV}$ the bound from the CMB would still marginally improve over existing constraints.

\vskip4pt
Massive axions are unstable to decay mediated by the operator $\phi F \tilde F$. However, for the range of parameters of interest, these decays occur after recombination and, hence, do not affect the CMB. To see this, we consider the decay rate for $m_\phi \gtrsim T$~\cite{Peskin:1995ev}, 
\beq
\Gamma_{D,\gamma} = \frac{1}{64\pi} \frac{m_\phi^3}{\Lambda_\gamma^2} \, .
\eeq
The decay time is $\tau_D = \Gamma_{D,\gamma}^{-1}$ and the temperature at decay is determined by $H(\Td) \approx \tau_D^{-1} = \Gamma_{D,\gamma}$. We will not consider the regime $m_\phi < \Td$ as it does not arise in the range of parameters of interest. Assuming that the universe is matter dominated at the time of the decay, we get
\beq
\frac{\Td}{\Trec} \,\approx\, \num{9.5e-10} \left(\frac{\Lambda_\gamma}{\SI{e10}{GeV}} \right)^{\!-4/3} \left(\frac{m_\phi}{\Trec}\right)^{\!2} \, .
\eeq
Using the stellar cooling constraint, $\Lambda_\gamma > \SI{1.3e10}{GeV}$~\cite{Friedland:2012hj}, we therefore infer that $\Td < \num{7.1e-10} \,\Trec \left({m_\phi/\Trec}\right)^{2}$, so that the axions are stable on the time-scale of recombination as long as $m_\phi \lesssim \SI{10}{keV}$.  CMB-S4 will probe this regime through sensitivity to $\Neff$ for $m_\phi  \lesssim \Trec$ and through sensitivity to warm dark matter for larger masses.  Warm dark matter is already highly constrained by cosmology, with current CMB data limiting the mass of the QCD axion to $m_\phi < \SI{0.53}{eV}$ (95\%\,C.L.)~\cite{DiValentino:2015wba}.  The regime $\SI{10}{keV} < m_\phi < \SI{1}{MeV}$ (where the axion decays between neutrino decoupling and recombination) is constrained by effects on the CMB and on Big Bang Nucleosynthesis (BBN)~\cite{Cadamuro:2010cz,Cadamuro:2011fd, Millea:2015qra}.

%%%%%%%%%%%%%%%%%%%%
\subsection{Coupling to Gluons}
\label{ssec:gluons}
%%%%%%%%%%%%%%%%%%%% 

\begin{figure}[t!]
\begin{center}
\includegraphics[width=0.8\textwidth]{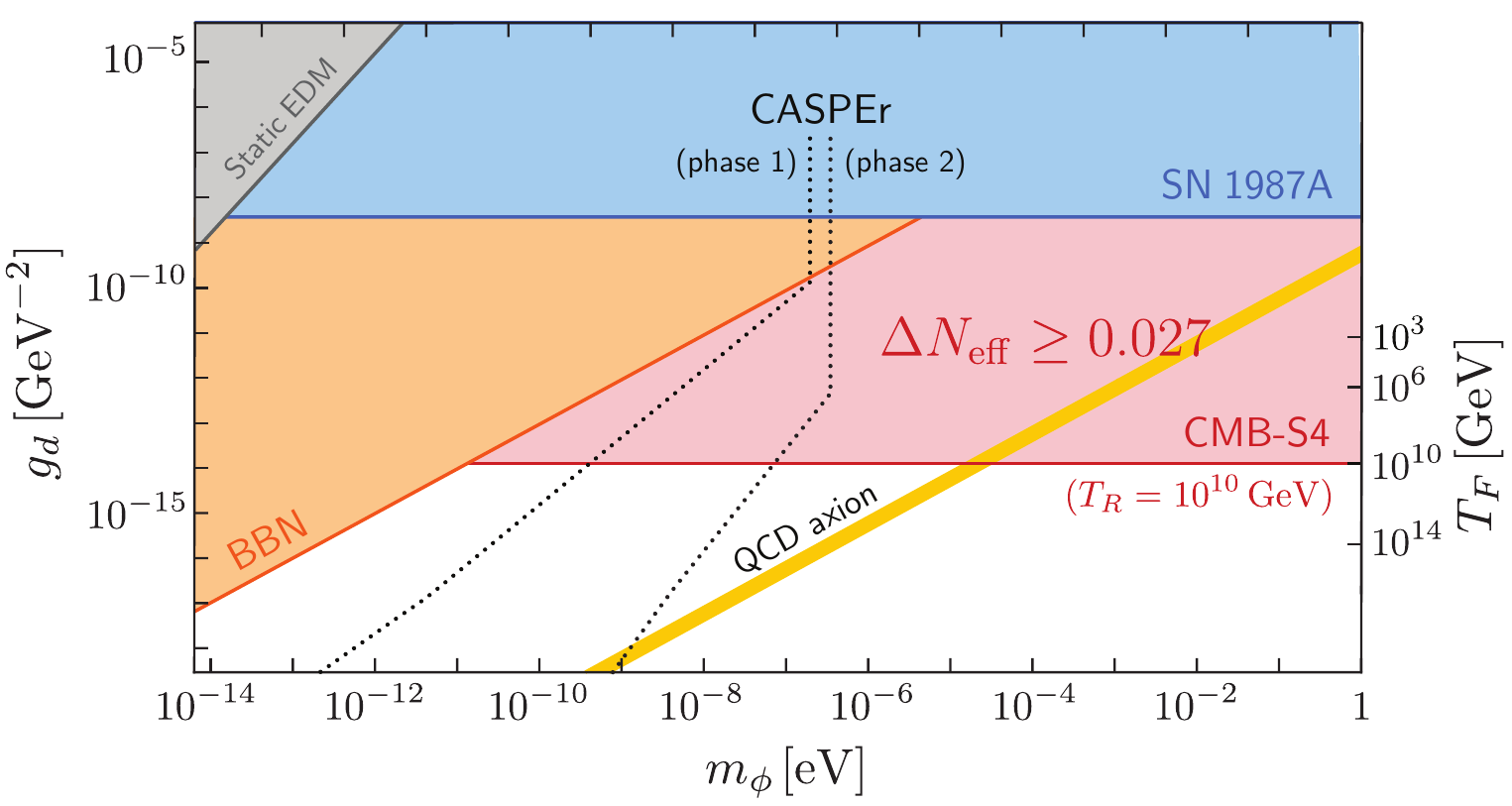}
\caption{Comparison between current constraints on the axion-gluon coupling and the sensitivity of a future CMB-S4 mission (figure adapted from~\cite{Graham:2013gfa,Blum:2014vsa}). The dotted lines are the projected sensitivities of the NMR experiment CASPEr~\cite{Budker:2013hfa}. We note that CASPEr, the static EDM~\cite{Graham:2013gfa} and BBN constraints~\cite{Blum:2014vsa} assume that the axion is all of the dark matter, while SN~1987A~\cite{Raffelt:1996wa} and the future CMB constraint do not have this restriction.}
\label{fig:S4dipole}
\end{center}
\end{figure}

The coupling to gluons is especially interesting for the QCD axion since it has to be present in order to solve the strong CP problem. The axion production rate associated with the interaction $\phi \hskip1pt G \tilde G$ is~\cite{Salvio:2013iaa}
\beq
\Gamma_g \simeq 0.41\hskip1pt \frac{T^3}{\Lambda_g^2}\, ,
\eeq
where $\Lambda_g \equiv \Lambda/c_3$. As before, we have dropped a weakly temperature-dependent prefactor, but account for it in Appendix~\ref{app:rates}. The bound~(\ref{eq:freezeoutconstraint}) then implies
\beq
\Lambda_g > \SI{5.4e13}{GeV}  \left(\frac{\Tr}{\SI{e10}{GeV}}\right)^{\!1/2}\, .
\eeq
Laboratory constraints on the axion-gluon coupling are usually phrased in terms of the induced electric dipole moment (EDM) of nucleons: $d_n = g_d \phi_0$, where $\phi_0$ is the value of the local axion field. The coupling $g_d$ is given for the QCD axion with an uncertainty of about 40\% by~\cite{Pospelov:1999ha, Graham:2013gfa}
\beq
g_d \,\approx\, \frac{2\pi}{\alpha_s} \times \frac{\SI{3.8e-3}{\per\GeV}}{\Lambda_g}  \ <\ \SI{1.3e-14}{\per\GeV\squared} \left(\frac{\Tr}{\SI{e10}{GeV}}\right)^{\!-1/2}\, .
\eeq
Constraints on $g_d$ (and hence $\Lambda_g$) are shown in Fig.~\ref{fig:S4dipole}. We see that future CMB-S4 observations will improve over existing constraints on $\Lambda_g$ by up to six orders of magnitude if $\Tr = \O(\SI{e10}{GeV})$. Even if the reheating temperature is as low as \SI{e4}{GeV}, the future CMB constraints will be tighter by  three orders of magnitude.  In Fig.~\ref{fig:S4dipole}, we also show the projected sensitivities of the proposed EDM experiment CASPEr~\cite{Budker:2013hfa}.  We see that CASPEr and CMB-S4 probe complementary ranges of axion masses.  It should be noted that CASPEr is only sensitive to axion dark matter, while the CMB constrains a separate thermal population of axions which does not require assumptions about the dark matter.

%%%%%%%%%%%%%%%%%%%
\section{Constraints on Familons}
\label{sec:fermions}
%%%%%%%%%%%%%%%%%%%

Spontaneously broken global symmetries have also been envoked to explain the approximate $U(3)^5$ flavor symmetry of the Standard Model. The associated pNGBs---called {\it familons}~\cite{PhysRevLett.49.1549,reiss1982can,kim1987light}---couple to the SM through Yukawa couplings,
\begin{align}
\L_{\phi\psi} &= - \frac{\partial_\mu \phi}{\Lambda_\psi}\, \bar \psi_i \gamma^\mu \big(g^{ij}_V+  g^{ij}_A \gamma^5\big) \psi_j \nonumber \\[4pt] &\to \frac{\phi}{\Lambda_\psi}\bigg( \i  H  \,\bar\psi_{L,i} \!\left[ (\lambda_i - \lambda_j) g_V^{ij} + (\lambda_i + \lambda_j) g_A^{ij}  \right]\! \psi_{R,j} + {\rm h.c.} \bigg) \,+\, \O(\phi^2)\, , \label{equ:Lfamilon}
\end{align}
where $H$ is the Higgs doublet and $\psi_{L,R} \equiv \tfrac{1}{2}(1\mp \gamma^5) \psi$.  The $SU(2)_L$ and $SU(3)_c$ structures  in~(\ref{equ:Lfamilon}) take the same form as for the SM Yukawa couplings~\cite{Peskin:1995ev}, but this has been left implicit to avoid clutter. In the second line we have integrated by parts and used the equations of motion.  The subscripts $V$ and $A$ denote the couplings to the vector and axial-vector currents, respectively, and $\lambda_i \equiv  \sqrt{2} m_i /v$ are the Yukawa couplings, with  $v = \SI{246}{GeV}$ being the Higgs vacuum expectation value.  We note that the diagonal couplings, $i=j$, are only to the axial part, as expected from vector current conservation. Due to the chiral anomaly, a linear combination of the axial couplings is equivalent to the coupling of axions to gauge bosons.  In this section, we only consider the effects of the couplings to matter with no contribution from anomalies.

\vskip4pt
In Table~\ref{tab:bounds}, we have collected accelerator and astrophysics constraints on the effective couplings $\Lambda_{ij}^I \equiv \Lambda_\psi/g^{ij}_I$ and $\Lambda_{ij} \equiv {\Lambda_\psi/[(g_V^{ij})^2 + (g_A^{ij})^2}]^{1/2}$.  We see that current data typically constrain the couplings to the first generation fermions much more than those to the second and third generations.  We wish to compare these constraints to the reach of future CMB observations.  We will find distinct behavior above and below the EWSB scale, due to the presence of the Higgs.  The effective scaling of the operator~(\ref{equ:Lfamilon}) changes from irrelevant to marginal and we therefore have both freeze-out and freeze-in contributions.
\begin{table}[t!]
\begin{center}
 \begin{tabular}{c c r c c c c} 
  \toprule
 						& \multicolumn{2}{c}{Current Constraints}					& & \multicolumn{3}{c}{Future CMB Constraints}	\\
  \cmidrule(lr){2-3} \cmidrule(lr){5-7}
	Coupling  			& Bound [GeV]		& Origin 							& & Freeze-Out [GeV]		& Freeze-In [GeV]		& $\Delta\tilde{N}_\mathrm{eff}$\\
  \midrule[0.065em]
	$\Lambda_{ee}$		& \ \num{1.2e10}		& White dwarfs 					& & \num{6.0e7}			& \num{2.7e6}		& 1.3\phantom{0}	\\[2pt] 
	$\Lambda_{\mu\mu}$	& \num{2.0e6}		& Stellar cooling					& & \ \num{1.2e10}		& \num{3.4e7}		& 0.5\phantom{0}	\\[2pt] 
	$\Lambda_{\tau\tau}$	& \num{2.5e4}		& Stellar cooling					& & \ \num{2.1e11}		& \num{9.5e7}		& 0.05			\\[2pt] 
	$\Lambda_{bb}$		& \num{6.1e5}		& Stellar cooling					& & \ \num{9.5e11}		& {--}	& 0.04			\\[2pt]
	$\Lambda_{tt}$		& \num{1.2e9}		& Stellar cooling					& & \ \num{3.5e13}		& {--}	& 0.03			\\
  \midrule
	$\Lambda_{\mu e}^V$	& \num{5.5e9}		& $\mu^+ \to e^+ \,\phi $			& & \num{6.2e9}			& \num{4.8e7}		& 0.5\phantom{0}	\\[2pt] 
	$\Lambda_{\mu e}$	& \num{3.1e9}		& $\mu^+ \to e^+ \,\phi\,\gamma$	& & \num{6.2e9}			& \num{4.8e7}		& 0.5\phantom{0}	\\[2pt] 
	$\Lambda_{\tau e}$	& \num{4.4e6}		& $\tau^- \to e^- \phi$ 				& & \ \num{1.0e11}		& \num{1.3e8}		& 0.05			\\[2pt] 
	$\Lambda_{\tau \mu}$	& \num{3.2e6}		& $\tau^- \to \mu^- \phi$ 			& & \ \num{1.0e11}		& \num{1.3e8}		& 0.05			\\[2pt] 
	$\Lambda_{cu}^A$	& \num{6.9e5}		& $D^{0}$-$\bar D^{0}$ 			& & \ \num{1.3e11}		& \num{2.0e8}		& 0.05			\\[2pt] 
	$\Lambda_{bd}^A$	& \num{6.4e5}		& $B^{0}$-$\bar B^{0}$ 			& & \ \num{4.8e11}		& \num{3.7e8}		& 0.04			\\[2pt] 
	$\Lambda_{bs}$		& \num{6.1e7}		& $b \to s \phi$					& & \ \num{4.8e11}		& \num{3.7e8}		& 0.04			\\[2pt] 
	$\Lambda_{tu}$		& \num{6.6e9}		& Mixing							& & \ \num{1.8e13}		& \num{2.1e9}		& 0.03			\\[2pt] 
	$\Lambda_{tc}$		& \num{2.2e9}		& Mixing							& & \ \num{1.8e13}		& \num{2.1e9}		& 0.03			\\[2pt] 
  \bottomrule 
 \end{tabular}
\caption{Current experimental constraints on Goldstone-fermion couplings (taken from~\cite{Brust:2013xpv, Feng:1997tn, Hansen:2015lqa}) and future CMB constraints. In some cases the current constraints are only on the coupling to right-handed particles (namely for $\Lambda_{\tau\tau}$, $\Lambda_{bb}$, $\Lambda_{tt}$) and to left-handed particles (namely for $\Lambda_{tu}, \Lambda_{tc}$).  The quoted freeze-out bounds are for $\Tr = \SI{e10}{GeV}$ and require that a future CMB experiment excludes $\Delta \Neff=0.027$.  In contrast, the freeze-in bounds from avoiding recoupling of the familons to the SM at low temperatures do not depend on $T_R$ and assume weaker exclusions $\Delta \tilde N_{\rm eff}$ [see the last column for estimates of the freeze-in contributions associated with the different couplings, $\Delta \tilde N_{\rm eff} \simeq \Delta N_{\rm eff}(\frac{1}{4}m_i)$]. Hence, they may be detectable with a less sensitive experiment. Qualitatively, the bounds from the CMB are stronger for the second and third generations, while laboratory and stellar constraints are strongest for the first generation (with the exception of the constraint on $\Lambda_{tt}$).}
\label{tab:bounds}
\end{center}
\end{table}

\subsection{Freeze-Out}

At high energies, the flavor structure of~(\ref{equ:Lfamilon}) is unimportant since all SM particles are effectively massless.  The role of the flavor is only to establish the strength of the interaction by the size of the Yukawa coupling.  Above the EWSB scale, the production of the familon $\phi$ is determined by a four-point interaction. This allows the following processes: $\bar\psi_i + \psi_j \rightarrow H +\phi$ and $\psi_i + H \rightarrow \psi_j + \phi$. The total production rate is derived in Appendix~\ref{app:rates},
\beq
\Gamma_{ij}^{I} \simeq 0.37 \hskip1pt N_\psi\, \frac{(\lambda_i \mp \lambda_j)^2}{8\pi} \frac{T^3}{(\Lambda_{ij}^{I})^2} \, ,
\label{equ:GijI}
\eeq
where $N_\psi=1$ for charged leptons and $N_\psi=3$ for quarks. The `$-$' and `$+$' signs in~(\ref{equ:GijI}) apply to $I=V$ and $I=A$, respectively.  We see that the rate vanishes for the diagonal vector coupling, as it should by current conservation. Deriving the freeze-out temperature and imposing $\Tf > \Tr$, we find 
\beq
\Lambda_{ij}^{I} \ >\ \left\{ \begin{array}{ll} \displaystyle \SI{1.0e11}{GeV} \ \frac{m_i \mp m_j}{m_\tau} \left(\frac{\Tr}{\SI{e10}{GeV}}\right)^{\!1/2} & \quad i,j=\text{leptons}, \\[10pt]
\displaystyle \SI{1.8e13}{GeV} \ \frac{m_i \mp m_j}{\mt} \left(\frac{\Tr}{\SI{e10}{GeV}}\right)^{\!1/2} & \quad i,j=\text{quarks},
\end{array}
\right.
\eeq
where $m_\tau \approx \SI{1.8}{GeV}$ and $\mt \approx \SI{173}{GeV}$. In Table~\ref{tab:bounds}, we show how these bounds compare to current laboratory and astrophysics constraints for a fiducial reheating temperature of $\SI{e10}{GeV}$. Except for the coupling to electrons, the constraints from future CMB experiments are orders of magnitude stronger than existing constraints.For lower reheating temperatures the constraints would weaken proportional to $\sqrt{\Tr}$. We note that while laboratory and astrophysical constraints are considerably weaker for second and third generation particles because of kinematics, the cosmological constraints are strengthened for the higher mass fermions due to the larger effective strength of the interactions.  The exception to this pattern is the top quark which is strongly constrained by stellar cooling due to a loop correction to the coupling of $W^\pm$ and $Z$ to~$\phi$, with the loop factor suppression being offset by the large Yukawa coupling of the top quark.

\subsection{Freeze-In}

Below the EWSB scale, the leading coupling of the familon to fermions becomes marginal after replacing the Higgs in~(\ref{equ:Lfamilon}) with its vacuum expectation value.  As the temperature decreases, the production rate will therefore grow relative to the expansion rate and we may get a thermal freeze-in abundance.  The leading familon production mechanism will depend on whether the coupling is diagonal or off-diagonal in the mass eigenbasis. 

\vskip4pt \noindent
{\it Diagonal couplings.}---For the diagonal couplings in~(\ref{equ:Lfamilon}), the production rate is dominated by a Compton-like process, $\{\gamma,g\} + \psi_i \to \psi_i + \phi$, and by fermion/anti-fermion annihilation,  $\bar \psi_i + \psi_i \to \{\gamma,g\} + \phi$, where $\{\gamma,g\}$ is either a photon or gluon depending on whether the fermion is a lepton or quark. The full expression for the corresponding production rate is given in Appendix~\ref{app:rates}.  Since freeze-in occurs at low temperatures, the quark production becomes sensitive to strong coupling effects.  Although qualitative bounds could still be derived for the quark couplings, we choose not to present them and instead focus on the quantitative bounds for the lepton couplings.  Below the scale of EWSB, but above the lepton mass, the production rate is
\beq
\tilde \Gamma_{ii} \simeq 5.3 \hskip1pt \alpha \, \frac{|\tilde \epsilon_{ii}|^2}{8 \pi} T\,  , 
\eeq
where $\tilde \epsilon_{ii} \equiv 2 m_i/\Lambda_{ii}$.  The freeze-in temperature $\Tff$ follows from $\tilde \Gamma_{ii}(\Tff) = H(\Tff)$. To avoid producing a large familon abundance requires that the fermion abundance becomes Boltzmann suppressed before freeze-in could occur. This implies $\Tff < m_i$, or 
\beq
 \Lambda_{ii} \ >\  \SI{9.5e7}{GeV} \left(\frac{g_{*,i}}{g_{*,\tau}}\right)^{\!-1/4} \left(\frac{\alpha_{i}}{\alpha_\tau}\right)^{\!1/2} \left(\frac{m_i}{m_\tau}\right)^{\!1/2}\, ,  \quad i=\text{lepton},  \label{equ:Lii}
\eeq
where $g_{*,i}$ and $\alpha_i$ are the effective number of relativistic species and the fine-structure constant at $T=m_i$. The scalings in~(\ref{equ:Lii}) have been normalized with respect to $g_*$ and $\alpha$ at $T= m_\tau$, i.e.~we use $g_{*,\tau} = 81.0$  and $\alpha_\tau = 134^{-1}$. In Table~\ref{tab:bounds}, these bounds are compared to the current astrophysical constraints. Except for the coupling to electrons, these new bounds are significantly stronger than the existing constraints. 

\vskip4pt \noindent
{\it Off-diagonal couplings.}---For the off-diagonal couplings in~(\ref{equ:Lfamilon}), we have the possibility of a freeze-in population of the familon from the decay of the heavy fermion, $\psi_i \to \psi_j +\phi$. For $m_i \gg m_j$, the production rate associated with this process is
\beq
\tilde\Gamma_{ij} \simeq 0.31\hskip1pt N_\psi\,\frac{|\tilde \epsilon_{ij}|^2}{8\pi} \frac{m_i^2}{T} \, ,
\eeq
where $\tilde \epsilon_{ij} \approx m_i/\Lambda_{ij}$. Requiring the corresponding freeze-in temperature to be below the mass of the heavier fermion, $\Tff < m_i$, we get 
\beq
\Lambda_{ij}  \ >\ \left\{ \begin{array}{ll} \displaystyle\SI{1.3e8}{GeV} \left(\frac{g_{*,i}}{g_{*,\tau}}\right)^{\!-1/4} \left(\frac{m_i}{m_\tau}\right)^{\!1/2} & \quad i,j=\text{leptons}, \\[10pt]
\displaystyle \SI{2.1e9}{GeV} \left(\frac{g_{*,i}}{g_{*,t}}\right)^{\!-1/4} \left(\frac{m_i}{m_t}\right)^{\!1/2} & \quad i,j=\text{quarks}.
\end{array} \right.
\eeq
We see that this improves over existing constraints for the third generation leptons and for the second and third generation quarks (except the top).

\vskip4pt
The freeze-in abundance is created {\it after} the annihilation of the most massive fermion in the coupling.  In the presence of a single massive fermion, the prediction for a freeze-in scenario is the same as that for a freeze-out scenario with $\Tf \ll m_i$ since decoupling occurs after most of the fermions $\psi_i$ have annihilated and their abundance is exponentially suppressed. This then results in a relatively large contribution to $\Neff$. Of course, the SM contains fermions with different masses.  To capture the energy injection from the relevant fermion annihilation without incorrectly including the effects from the annihilation of much lighter fermions, we take the decoupling temperature to be $\tfrac{1}{4}m_i$.  This choice of decoupling temperature gives good agreement with numerical solutions to the Boltzmann equations and leads to the following
estimate for the freeze-in contributions:
\beq
\Delta \tilde N_{\rm eff} \simeq \Delta\Neff(\tfrac{1}{4}m_i)= \frac{4}{7} \left(\frac{43}{4\,g_*(\tfrac{1}{4}m_i)}\right)^{4/3}\, .
\eeq
When the heaviest fermion is a muon (electron), one finds $\Delta \tilde N_{\rm eff}\simeq 0.5$ ($1.3$) which is excluded by Planck at about $3\sigma$ ($7\sigma$).  It is worth noting that the Planck constraint on the diagonal muon coupling, $\Lambda_{\mu\mu}>\SI{3.4e7}{GeV}$, improves on the current experimental bound by more than an order of magnitude.  Couplings involving the tau and the charm or bottom quark produce $\Delta \tilde N_{\rm eff} \sim 0.05$ which will become accessible when the sensitivity of CMB experiments reaches $\sigma(\Neff) \lesssim 0.025$.

%%%%%%%%%%%%%%%%%%%%%
\section{Constraints on Majorons}
\label{sec:neutrinos}
%%%%%%%%%%%%%%%%%%%%%

In the Standard Model, the masses of Majorana neutrinos do not arise from renormalizable couplings to the Higgs, but instead must be written as irrelevant operators suppressed by a scale of about \SI{e15}{GeV}.  Moreover, the existence of neutrino masses and mixings point to structure in the flavor physics of neutrinos.  Much like in the case of familons, it is plausible that this structure could arise from the spontaneous breaking of the neutrino flavor symmetry. The Goldstone bosons associated with this SSB are often referred to as {\it majorons}~\cite{Chikashige:1980ui, Chikashige:1980qk}.  

\vskip4pt
Assuming that neutrinos are indeed Majorana fermions, the leading coupling of the majoron~is
\begin{align}
\L_{\phi\nu} &= - \frac{1}{2} \left(e^{\i \phi T_{ik}/ (2\Lambda_\nu)} \hskip1pt m_{kl} \hskip1pt e^{\i \phi T_{l j}/(2 \Lambda_\nu)} \nu_i \nu_j +{\rm h.c.}\right)  \nonumber \\[4pt]
&= - \frac{1}{2}\left[ \left(m_{ij} \nu_i \nu_j  + \i \hskip1pt \tilde \epsilon_{ij} \phi \nu_i \nu_j - \frac{1}{2\Lambda_\nu} \epsilon_{ij}  \phi^2 \nu_i \nu_j + \cdots \right) + {\rm h.c.}\right] ,  \label{equ:Lmajoron} 
\end{align}
where $\nu_{i}$ are the two-component Majorana neutrinos in the mass eigenbasis, $m_{ij}$ is the neutrino mass matrix and $T_{ij}$ are generators of the neutrino flavor symmetry.  After expanding the exponentials, we have defined the dimensionless couplings $\tilde \epsilon_{ij} \equiv  (T_{i k} m_{k j} +m_{i k} T_{k j})/(2\Lambda_\nu)$ and $\epsilon_{ij} \equiv (m_{ik} T_{kl} T_{lj} + 2 T_{ik} m_{kl} T_{lj} +T_{ik} T_{kl} m_{lj})/(4\Lambda_\nu)$.  For numerical estimates,
we will use the cosmological upper limit on the sum of the neutrino masses~\cite{Ade:2015xua}, $\sum m_i < \SI{0.23}{eV}$, and the mass splittings $m_2^2-m_1^2 \approx \SI{7.5e-5}{eV\squared}$ and $|m_3^2 - m_1^2| \approx \SI{2.4e-3}{eV\squared}$ from neutrino oscillation measurements~\cite{Agashe:2014kda}.  The couplings in $\L_{\phi\nu}$ are identical to the familon couplings after a chiral rotation, except that there is no analogue of the vector current in the case of Majorana neutrinos.  The representation of the coupling  in~(\ref{equ:Lmajoron}) is particularly useful as it makes manifest both the marginal and irrelevant couplings between $\phi$ and $\nu$.  As a result, we will get both a freeze-out\hskip1pt\footnote{Technically speaking the operator in~(\ref{equ:Lmajoron}) is only well-defined below the EWSB scale. However, in~\S\ref{sec:nufreeze} we will find  that in order for freeze-out to occur in the regime of a consistent effective field theory description ($T< \Lambda_\nu$), we require $\Tf \lesssim \SI{33}{MeV}$ and, therefore, the operator as written will be sufficient for our purposes.} and a freeze-in production of the majorons.

\subsection{Freeze-Out}
\label{sec:nufreeze}

Thermalization at high energies is dominated by the dimension-five operator $\phi^2 \nu_i \nu_j$ in~(\ref{equ:Lmajoron}).  In Appendix~\ref{app:rates}, we show that the corresponding production rate is 
\beq
\Gamma_{ij} \simeq 0.047 \hskip1pt s_{ij} \, \frac{| \epsilon_{ij}|^{2}}{8\pi} \,\frac{T^3}{\Lambda_\nu^2} \, , 
\eeq
where $s_{ij} \equiv 1-\frac{1}{2} \delta_{ij}$ is the symmetry factor for identical particles in the initial state.  This leads to a freeze-out temperature of
\beq
 \Tf \simeq \SI{0.23}{MeV}  \,  s_{ij}^{-1}  \, \bigg(\frac{g_{*,F}}{10}\bigg)^{\!1/2} \left(\frac{\mu_{ij}}{\SI{0.1}{eV}}\right)^{\!-2} \left(\frac{\Lambda_\nu}{\SI{10}{MeV}}\right)^{\!4} \, , \label{equ:Tffam}
\eeq
where $\mu_{ij} \equiv |\epsilon_{ij}| \Lambda_\nu$. Consistency of the effective field theory (EFT) description requires $\Tf$ to be below the cutoff $\Lambda_\nu$ associated with the interactions in~(\ref{equ:Lmajoron}). Using~(\ref{equ:Tffam}), this implies 
\beq
\Tf < \Lambda_\nu <\, \SI{35}{MeV} \, s_{ij}^{1/3}   \left(\frac{g_{*,F}}{10}\right)^{\!-1/6} \left(\frac{\mu_{ij}}{\SI{0.1}{eV}}\right)^{\!2/3}\, .
\eeq
Taking $\mu_{ij} \lesssim m_3<\SI{0.1}{eV}$ from both the mass splittings and the bound on the sum of neutrino masses and $g_* \approx 14$, we obtain $\Tf\lesssim \SI{33}{MeV}$.  Such a low freeze-out temperature would lead to $\Delta\Neff\gtrsim0.44$ (cf.~Fig.~\ref{fig:freezeout}) which is ruled out by {\it current} CMB measurements at more than $2\sigma$.  To avoid this conclusion, we require $\Lambda_\nu > \SI{33}{MeV}$, so that the would-be freeze-out is pushed outside the regime of validity of the EFT.  Moreover, we have to assume that the production of majorons is suppressed in this regime.  This logic leads to the following constraint: 
\beq
\Lambda_\nu > \SI{33}{MeV} \quad \xrightarrow{\ \mu_{ij} \,\lesssim\, \SI{0.1}{eV} \ } \quad  |\epsilon_{ij}| < \num{3e-9} \, . \label{eqn:nufreezeout}
\eeq
Somewhat stronger bounds can be derived for individual elements of $\epsilon_{ij}$. This simple bound is much stronger than existing constraints from neutrinoless double beta decay~\cite{Gando:2012pj, Albert:2014fya} and supernova cooling~\cite{Farzan:2002wx}, $\epsilon_{ij} \lesssim 10^{-7}$.  Note also that the constraints on $\epsilon_{ij}$ are stronger for smaller values of $\mu_{ij}$.

\subsection{Freeze-In}
\label{sec:nufreezein}

At low energies, the linear coupling $\phi \nu_i  \nu_j$ in~(\ref{equ:Lmajoron}) will dominate.  The corresponding two-to-one process is kinematically constrained and we therefore get qualitatively different behavior depending on whether the majoron mass is larger or smaller than that of the neutrinos.

\vskip4pt \noindent
{\it Low-mass regime.}---For $m_\phi \ll m_i -m_j$, with $m_i > m_j$, the off-diagonal couplings allow the decay $\nu_i \to \nu_j + \phi$, while other decays are kinematically forbidden.  As a result, only the off-diagonal couplings are constrained by freeze-in.  Including the effect of time dilation at finite temperature, the rate is
\beq
\tilde\Gamma_{ij} \simeq 0.31\,  \frac{|\tilde \epsilon_{ij}|^2}{8\pi}\, \frac{m_i^2}{T} \, , \label{eqn:nudecay}
\eeq
where we have assumed $m_i \gg m_j$, which is guaranteed for the minimal mass normal hierarchy (for the general result see~Appendix~\ref{app:rates}).  When the freeze-in occurs at $\Tff>m_i$, then the majorons and neutrinos are brought into thermal equilibrium, while the comoving energy density is conserved. However, since the momentum exchange at each collision is only $\Delta p^2 \simeq m_i^2 \ll T^2$, the neutrino-majoron radiation is free-streaming at the onset of the freeze-in and is difficult\footnote{Since neutrinos have been converted to majorons with $m_\phi \ll m_i$, this scenario predicts that the cosmological measurement of the sum of the neutrino masses would be significantly lower than what would be inferred from laboratory measurements.} to distinguish from conventional neutrinos.  As the temperature drops below $\Tfl$, with $\tilde\Gamma_{ij}(\Tfl) = (\Tfl/m_i)^2\, H(\Tfl)$, enough momentum is exchanged between the neutrinos and they will behave as a relativistic fluid rather than free-streaming particles~\cite{Chacko:2003dt, Hannestad:2005ex, Friedland:2007vv}.  From the rate~(\ref{eqn:nudecay}), we find
\beq
\Tfl \simeq \\ \num{0.10} \, \Teq \times \left(\frac{\tilde \epsilon_{ij}}{10^{-13}} \right)^{\!2/5} \,  \left(\frac{m_i}{\SI{0.05}{eV}} \right)^{\!4/5}  \, , 
\eeq
where we used $g_{*,\tilde F} \approx 3.4$ and $\Teq\approx\SI{0.79}{eV}$ for the temperature at matter-radiation equality.  In analogy with~(\ref{equ:Neff}), we write the energy density of the fluid as $\Delta\rho_r \equiv \frac{7}{8}(\tfrac{4}{11})^{4/3}\, N_{\rm fluid}\, \rho_\gamma$.  In this regime, the majoron scenario predicts  $N_{\rm fluid} \geq 1$ and $\Neff \leq 2$ (with equality when the majoron couples to only a single neutrino species), which is inconsistent with recent constraints from Planck data~\cite{Baumann:2015rya}: $\Neff=2.99 \pm 0.30 \hskip 3pt (68\%\,{\rm C.L.})$ and $N_{\rm fluid} < 1.06 \hskip 3pt (95\%\,{\rm C.L.})$. To avoid this conclusion requires $\Tfl < \Teq$,\footnote{The imprint of dark radiation is suppressed during matter domination since its contribution to the total energy density is subdominant.  As a result, constraints on $\Neff$ are driven by the high-$\ell$ modes of the CMB which are primarily affected by the evolution of fluctuations during radiation domination~\cite{Bashinsky:2003tk, Baumann:2015rya}.} which puts a bound on the neutrino-majoron coupling\hskip1pt\footnote{The effect of the linear coupling between a massless majoron and neutrinos on the CMB was also studied in~\cite{Forastieri:2015paa} and a flavor-independent constraint of $\tilde\epsilon_{ij} < \num{8.2e-7}$ was obtained.  This constraint is substantially weaker than our bound~(\ref{eqn:lowmassbound}) because it only accounted for the scattering of neutrinos through the exchange of a {\it virtual} Goldstone boson. The neutrino cross section in that case is suppressed by a factor of $|\tilde \epsilon_{ij}|^4$ which is much smaller than the rate for the production of real Goldstone bosons in~(\ref{eqn:nudecay}).} 
\beq
\tilde\epsilon_{ij} \,<\, \num{3.2e-11} \times \left(\frac{m_i}{\SI{0.05}{eV}}\right)^{\!-2}   \, . 	\label{eqn:lowmassbound}
\eeq
This constraint has been pointed out previously in~\cite{Chacko:2003dt, Hannestad:2005ex, Friedland:2007vv, Archidiacono:2013dua}.

\vskip4pt \noindent
{\it High-mass regime.}---For  $m_\phi \gg m_i \ge m_j$, the majoron decays into neutrinos, $\phi \to \nu_i+ \nu_j$, and is produced by the inverse decay.  For $T \gg m_\phi$, the production rate of the majoron is identical to the rate in~(\ref{eqn:nudecay}) after making the replacement $m_i \to m_\phi/\sqrt{1- 4/\pi^2}$ and the corresponding freeze-in temperature is
\beq
\Tff \simeq 1.0 \,\Teq \times \, s_{ij}^{1/3}   \left(\frac{\tilde \epsilon_{ij}}{10^{-13}} \,  \frac{m_\phi}{\Teq} \right)^{\!2/3} \, .
\eeq
If $\Tff>m_\phi$, then freeze-in occurs while the majorons are relativistic, and the neutrinos and the majorons are brought into thermal equilibrium.  How this affects the CMB will depend on whether $m_\phi$ is greater or smaller than $\Teq$. For $m_\phi > \Teq$, the majorons decay to neutrinos before matter-radiation equality. To compute the effect on the CMB, we note that the initial (relativistic) freeze-in process conserves the comoving energy density and, once in equilibrium, the decay will conserve the comoving entropy density. This information allows us to derive the final neutrino temperature analytically (see Appendix~\ref{app:decays}) and to determine the extra contribution to the radiation density,
\beq
\Delta \Neff \,\geq\, \left(1+ \frac{4}{7}\right)^{\!1/3} -1 =0.16\, . \label{equ:DNeffnu}
\eeq 
This extra radiation density is easily falsifiable (or detectable) with future CMB experiments.  If $m_\phi \ll \Teq$, on the other hand, the neutrinos and the majorons could effectively form a fluid at matter-radiation equality leading to a similar constraint as~(\ref{eqn:lowmassbound}) with $m_i \to m_\phi$.

Assuming that future experiments do not detect the above effects would require either that the would-be freeze-in temperature is below the mass of the majoron, $\Tff < m_\phi$, or that freeze-in occurs after matter-radiation equality, $\Tff < \Teq$.  Converting these constraints into a bound on the coupling, we find
\beq
\tilde \epsilon_{ij} \,<\, \num{9.9e-14} \,  s_{ij}^{-1/2}   \left(\frac{ m_\phi}{\Teq} \right)^{\!1/2}\, , \quad {\rm for} \quad m_\phi>\Teq\, .
\eeq 
A similar constraint, of the same order of magnitude, applies in the narrow range $m_i \ll m_\phi<\Teq$. This bound is stronger than the freeze-out constraint~(\ref{eqn:nufreezeout}) over the full range of allowed masses up to the neutrino decoupling temperature $T_{F,\hskip1pt\nu} \simeq \SI{1}{MeV}$ (note that although in general $\epsilon_{ij} \neq \tilde \epsilon_{ij}$, the two parameters are related by the symmetry under which the majoron transforms). For $m_\phi > T_{F,\hskip1pt\nu}$, the decay of the majorons occurs while the neutrinos are still in equilibrium with the SM and, therefore, it has no impact on $\Neff$.

%%%%%%%%%%%%%%%%%%%%%%
\section{Conclusions and Outlook}
\label{sec:conclusions}
%%%%%%%%%%%%%%%%%%%%%%

Light pseudo-Nambu-Goldstone bosons arise naturally in many proposals for physics beyond the Standard Model and are an exciting window into the early universe. In this paper, we showed that future CMB experiments will either detect these new particles, or place very strong constraints on their couplings to the SM. These constraints arise because the couplings to the SM can bring the Goldstone bosons into thermal equilibrium in the early universe. At the same time, cosmological experiments are becoming sensitive enough to detect thermal relics up to arbitrarily high freeze-out temperatures (see Fig.~\ref{fig:freezeout}).  To avoid producing this detectable relic abundance requires that the reheating temperature of the universe was below the would-be freeze-out temperature.  In that case, the temperature in the universe simply was never high enough to bring the extra particles into thermal equilibrium with the SM.  For a given reheating temperature $\Tr$, this puts bounds on the scales $\Lambda_i$ in the effective interactions between the Goldstone boson~$\phi$ and the SM fields,
\beq
\L_{\phi{\rm SM}} = - \frac{1}{4}\frac{\phi}{\Lambda_\gamma} F \tilde F  - \frac{1}{4}\frac{\phi}{\Lambda_g} {\rm Tr}(G \tilde G) -  \frac{\partial_\mu \phi}{\Lambda_\psi}\, \bar \psi \gamma^\mu  \gamma^5 \psi \, +\, \cdots\, .
\eeq 
The bounds on the couplings to photons and gluons are
\begin{align}
\Lambda_\gamma &> \SI{1.4e13}{GeV}\,\sqrt{T_{R,10}} \, , \label{eq:conclusionAxion}  \\
\Lambda_g &> \SI{5.4e13}{GeV}\,\sqrt{T_{R,10}} \, ,  
\end{align}
where $T_{R,10} \equiv \Tr/\SI{e10}{GeV}$.  When considering the interactions with fermions, we distinguish between the couplings to charged leptons and quarks. The resulting bounds are 
\begin{align}
\Lambda_\psi &\,>\, \left\{ \begin{array}{ll}\,  \SI{2.1e11}{GeV} \,\,m_{\psi,\tau}\,\sqrt{T_{R,10}}  & \quad \psi=\text{lepton}, \\[10pt] \, \SI{3.5e13}{GeV}\,\,m_{\psi,t}\,\sqrt{T_{R,10}}  & \quad \psi=\text{quark}, \end{array}\right. \label{eq:conclusionFamilon}
\end{align}
where $m_{\psi,\tau} \equiv m_\psi/\SI{1.8}{GeV}$ and $m_{\psi,t} \equiv m_\psi/\SI{173}{GeV}$.  For all reasonable reheating temperatures these bounds improve significantly over existing constraints, sometimes by many orders of magnitude. Moreover, while some of the current constraints only apply if the new particles are identified with the dark matter, our bounds do not have this restriction.

\vskip4pt
Below the scale of electroweak symmetry breaking, the couplings to the SM fermions become effectively marginal which can bring the decoupled Goldstone bosons back into thermal equilibrium leading to a detectable freeze-in abundance. Moreover, the coupling to the light Goldstone boson can lead to a new force between the fermions which becomes relevant at low temperatures~\cite{Chacko:2003dt, Hannestad:2005ex, Friedland:2007vv, Archidiacono:2013dua}. Both of these effects are highly constrained, even with current data~\cite{Baumann:2015rya}. These arguments are particularly relevant for the couplings to neutrinos, 
\beq
\L_{\phi\nu} = -\frac{1}{2}  \big(\,\i \hskip1pt \tilde\epsilon_{ij} \phi \nu_i \nu_j + \mathrm{h.c.}\,\big)+ \cdots\, . 
\eeq
For the off-diagonal couplings, the following constraints apply
\begin{align}
\tilde \epsilon_{ij} &\,<\ \left\{ \begin{array}{ll}\, \displaystyle \num{3.2e-11} \times \left(\frac{m_i}{\SI{0.05}{eV}}\right)^{\!-2}  & \quad m_\phi \ll m_i\, , \\[14pt] \, \displaystyle \num{9.9e-14}   \times  \left(\frac{ m_\phi}{\Teq} \right)^{\!1/2}  & \quad m_\phi > \Teq\, , \end{array}\right.
\end{align}
where $m_i$ is the mass of the heavier neutrino in the off-diagonal interaction.  A combination of freeze-in and freeze-out also constrain the diagonal couplings $\tilde \epsilon_{ii}$. 
These constraints are orders of magnitude stronger than existing laboratory and astrophysics constraints.

\vskip6pt
It is also interesting to consider a scenario in which one of the many ongoing searches directly detects axions, familons or majorons. This would determine the coupling strength to at least one of the SM fields (depending on the detection channel) and would predict the freeze-out temperature of these particles; cf.~Figs.~\ref{fig:S4axion} and~\ref{fig:S4dipole}. Excitingly, the cosmological estimation of $\Delta\Neff$ would then provide information about the reheating temperature of the universe: the absence of a detection of $\Neff \ne 3.046$ would put an upper bound on $\Tr$ [see e.g.~(\ref{eq:conclusionAxion})--(\ref{eq:conclusionFamilon})], while a measurement of $\Delta\Neff \geq 0.027$ would imply a lower bound on $\Tr$. The combination of a cosmological measurement of $\Neff$ and a direct detection could therefore be used to probe the energy scale of the beginning of the hot Big Bang.

\vskip6pt
In closing, we would like to re-emphasize that $\Delta \Neff = 0.027$ is an important theoretical threshold. Remarkably, this target is within reach of future cosmological observations~\cite{Font-Ribera:2013rwa}, including the planned CMB-S4 mission~\cite{Abazajian:2013oma}. These observations therefore have the potential to probe for light thermal relics up to arbitrarily high decoupling temperatures.  We consider this to be a unique opportunity to detect new particles, or place very strong constraints on their couplings to the Standard Model.

\vskip23pt
\paragraph{Acknowledgements}
We thank Jens Chluba, Nathaniel Craig, Daniel Grin, Julien Lesgourgues, David Marsh, Joel Meyers and Surjeet Rajendran for helpful discussions. D.\,G.~and B.\,W.~thank the Institute of Physics at the University of Amsterdam for its hospitality. D.\,B.~and B.\,W.~acknowledge support from a Starting Grant of the European Research Council (ERC STG Grant 279617). B.\,W.~is also supported by a Cambridge European Scholarship of the Cambridge Trust and an STFC Studentship. D.\,G.~was supported by an NSERC Discovery Grant and the Canadian Institute for Advanced Research.

\clearpage
\appendix
%%%%%%%%%%%%%%%%
\section{Production Rates}
\label{app:rates}
%%%%%%%%%%%%%%%%

In this appendix, we derive the rates of Goldstone boson production used in the main text.  We consider separately the couplings to gauge fields and to matter fields.

\subsection{Couplings to Gauge Fields}
\label{app:ratesGauge}
	
\begin{figure}[b]
\centering
\subfloat[\label{fig:gaugePrimakoffProcess}Primakoff process.]{
\includegraphics*{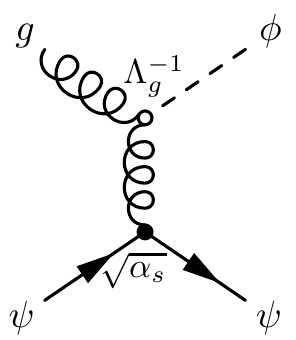}
}\hspace{0.5cm}
\subfloat[\label{fig:gaugeFermionAnnihilation}Fermion annihilation.]{
\includegraphics*{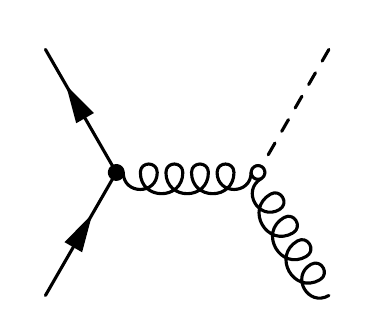}
}\hspace{0.5cm}
\subfloat[\label{fig:gaugeFusion}Gluon fusion (representative diagrams).]{
\includegraphics*{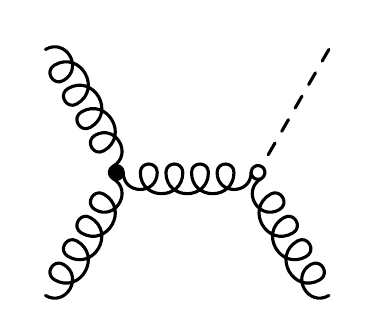}
\hspace{-0.4cm}
\includegraphics*{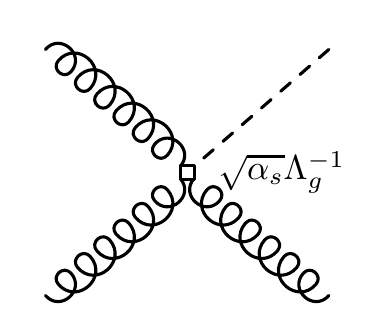}
}
\caption{Feynman diagrams for the dominant Goldstone production via the gluon coupling. For gluon fusion, there are $t$- and $u$-channel diagrams in addition to the presented $s$-channel diagram. Similar diagrams apply for the couplings to the electroweak gauge bosons.}
\label{fig:diagramsGluon}
\end{figure}

Above the scale of EWSB, the coupling of the Goldstone boson to the SM gauge sector is
\beq
\L_{\phi{\rm EW}} = - \frac{1}{4}\frac{\phi}{\Lambda} \left( c_1 \hskip1pt B_{\mu\nu} \tilde{B}^{\mu\nu} + c_2 \hskip1pt W_{\mu\nu}^a \tilde{W}^{\mu\nu,a} +c_3 \hskip1pt G_{\mu\nu}^a \tilde{G}^{\mu\nu,a} \right)  .  \label{equ:LEW}
\eeq
The dominant processes leading to the production of the Goldstone boson $\phi$ are illustrated in Fig.~\ref{fig:diagramsGluon}. In the limit of massless gauge bosons, the cross sections for some of these processes have infrared (IR) divergences.
The results therefore depend slightly on how these divergences are regulated; see e.g.~\cite{Braaten:1991dd, Bolz:2000fu, Masso:2002np, Graf:2010tv, Salvio:2013iaa}. The most detailed analysis has been performed in~\cite{Salvio:2013iaa} and the total production rate was found to be	
\beq
\Gamma = \frac{T^3}{8\pi \Lambda^2} \Big[ c_1^2\hskip1pt F_1(T) +  3 c_2^2 \hskip1pt F_2(T) +  8 c_3^2  \hskip1 pt F_3(T)  \Big]\, , \label{equ:TotalRate}
\eeq
where the functions $F_n(T)$ were derived numerically. We extracted $F_n(T)$ from Fig.~1 of~\cite{Salvio:2013iaa}, together with the one-loop running of the gauge couplings $\alpha_i(T)$.

\begin{figure}[t]
\centering
\includegraphics*{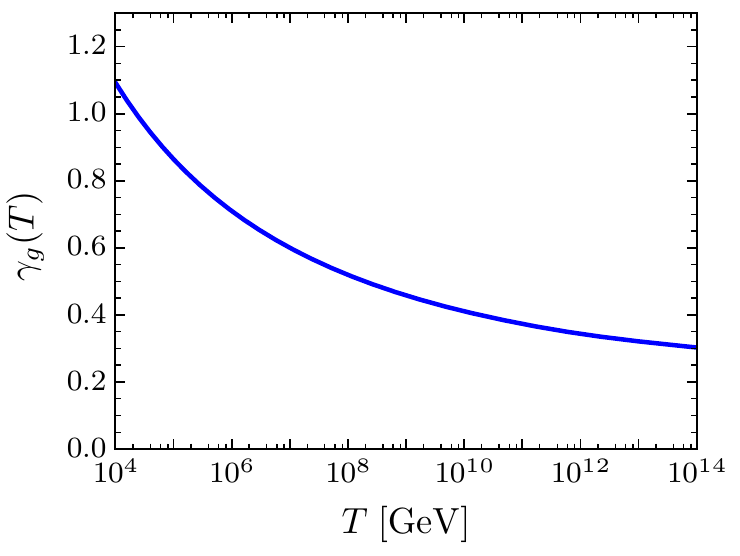}
\hspace{0.7cm}
\includegraphics*{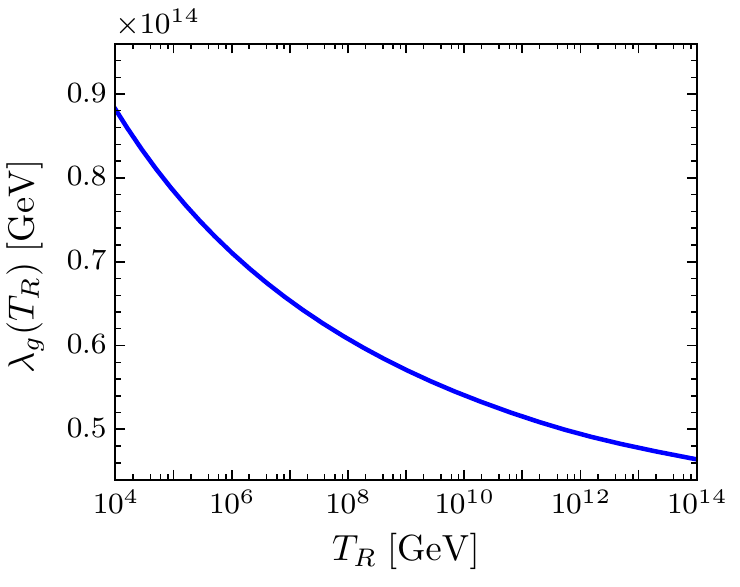}
\caption{{\it Left}: Axion production rate associated with the coupling to gluons as parametrized by~$\gamma_g(T)$ in~(\ref{equ:Gg}). {\it Right}: Constraint on the axion-gluon coupling $\Lambda_g$ as parametrized by $\lambda_g(T_R)$ in~(\ref{equ:Lambg}).}
\label{fig:rateGluon}
\end{figure}

\paragraph{Coupling to gluons}
To isolate the effect of the coupling to gluons, we write $c_1=c_2 \equiv 0$  and define $\Lambda_g \equiv \Lambda/c_3$. The production rate~(\ref{equ:TotalRate}) then becomes
\beq
\Gamma_g(T) = \frac{F_3(T)}{\pi} \frac{T^3}{\Lambda_g^2} \equiv \gamma_g(T) \frac{T^3}{\Lambda_g^2}\, , \label{equ:Gg}
\eeq
where $\gamma_g(\SI{e10}{GeV}) = 0.41$.  The function $\gamma_g(T)$ is presented in the left panel of Fig.~\ref{fig:rateGluon}.
The freeze-out bound on the gluon coupling then is
\beq
\Lambda_g \, >\, \left(\frac{\pi^2}{90} g_{*,R}\right)^{\!-1/4} \sqrt{\gamma_{g,R} \hskip1pt \Tr\hskip1pt \Mp} \,\, \equiv \, \lambda_g(\Tr) \left(\frac{\Tr}{\SI{e10}{GeV}}\right)^{\!1/2}\, , \label{equ:Lambg}
\eeq
where $g_{*,R} \equiv g_*(\Tr)$ and $\gamma_{g,R}\equiv \gamma_g(\Tr)$. The bound in~(\ref{equ:Lambg}) is illustrated in the right panel of Fig.~\ref{fig:rateGluon}. In the main text, we used $\lambda_g(\SI{e10}{GeV}) = \SI{5.4e13}{GeV}$.

\begin{figure}
\centering
\includegraphics*{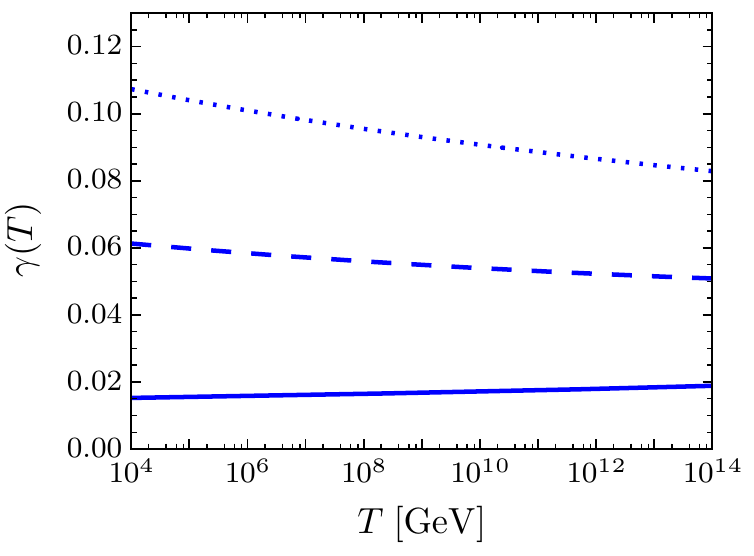}
\hspace{0.5cm}
\includegraphics*{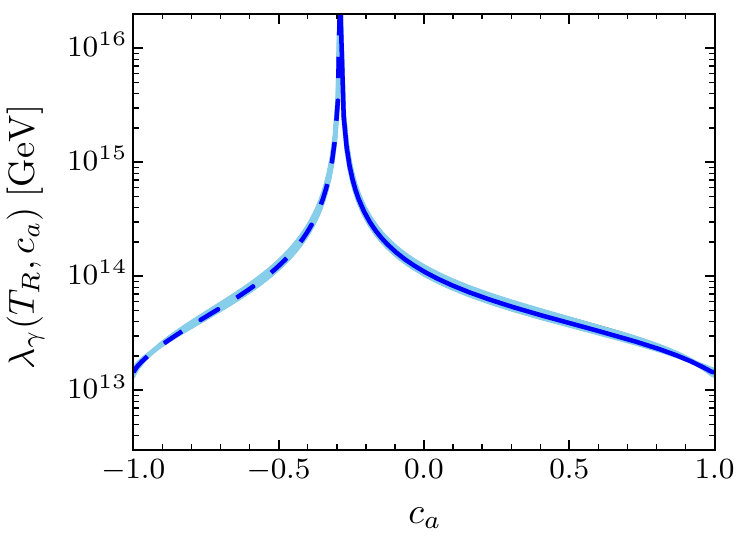}
\caption{{\it Left}: Axion production rate associated with the coupling to the electroweak gauge bosons as parametrized by $\gamma(T,c_a)$ in~(\ref{equ:GGgamma}) for $c_a=0$ (dotted line), $1/\sqrt{2}$ (dashed line) and 1 (solid line). {\it Right}:~Constraint on the axion-photon coupling $\Lambda_\gamma$ as parametrized by $\lambda_\gamma(\Tr, c_a)$ in~(\ref{equ:Lambgamma}).  The solid and dashed lines correspond to bounds on positive and negative $\Lambda_\gamma$ for $\Tr =\SI{e10}{GeV}$. The band displays the change for reheating temperatures between \SI{e4}{GeV} (upper edge) and \SI{e15}{GeV} (lower edge).}
\label{fig:ratePhoton}
\end{figure}

\paragraph{Coupling to photons}
To isolate the coupling to the electroweak sector, we set $c_3=0$.  In this case, the Lagrangian~(\ref{equ:LEW}) can be written as
\beq
\L_{\phi\rm EW} =  - \frac{1}{4}\frac{\phi}{\Lambda} \left( c_a \hskip1pt B_{\mu\nu} \tilde{B}^{\mu\nu} + s_a \hskip1pt W_{\mu\nu}^a \tilde{W}^{\mu\nu,a} \right)  , \label{equ:A5}
\eeq
where we have defined
\beq
\Lambda \to \frac{\Lambda}{\sqrt{c_1^2+c_2^2}} \quad {\rm and} \quad c_a \equiv \frac{c_1}{\sqrt{c_1^2 +c_2^2}} \ , \ \ s_a \equiv \frac{c_2}{\sqrt{c_1^2 +c_2^2}}\, .
\eeq
Note that $c_a^2 + s_a^2 =1$, so we can use $\Lambda$ and $c_a$ as the two free parameters. The production rate~(\ref{equ:TotalRate}) is then given by
\beq
\Gamma = \frac{[ c_a^2 F_1(T) +  3 s_a^2 \hskip1pt F_2(T) ]}{8\pi }  \frac{T^3}{\Lambda^2}\equiv \gamma(T,c_a)\hskip2pt \frac{T^3}{\Lambda^2}\, . \label{equ:GGgamma}
\eeq
The function $\gamma(T,c_a)$ is shown in the left panel of Fig.~\ref{fig:ratePhoton}. In the main text, we employed $\gamma(\SI{e10}{GeV},1) = 0.017$.
The freeze-out bound on the coupling then is
\beq
\Lambda(c_a) >  \left(\frac{\pi^2}{90} g_{*,R}\right)^{\!-1/4} \sqrt{\gamma_{R}(c_a) \hskip1pt \Tr\hskip1pt \Mp}\, , \label{equ:Lca}
\eeq
where $\gamma_R(c_a)\equiv \gamma(\Tr,c_a)$.  We wish to relate this bound to the couplings below the EWSB scale.

\vskip4pt
At low energies, the axion couplings to the electroweak sector become
\beq
\L_{\phi\rm EW} =- \frac{1}{4} \left( \frac{\phi}{\Lambda_\gamma} \hskip1pt F_{\mu \nu} \tilde{F}^{\mu \nu} + \frac{\phi}{\Lambda_{Z}}  Z_{\mu \nu} \tilde{Z}^{\mu \nu} + \frac{\phi}{\Lambda_{Z\gamma}} \hskip1pt Z_{\mu \nu} \tilde{F}^{\mu \nu} + \frac{\phi}{\Lambda_{W}}  W^+_{\mu \nu} \tilde{W}^{-}{}^{\mu \nu} \right)  ,  \label{equ:A8}
\eeq
where $F_{\mu \nu}$, $Z_{\mu \nu}$ and $W^{\pm}_{\mu\nu}$ are the field strengths for the photon, $Z$ and $W^\pm$, respectively.  Here, we have dropped additional (non-Abelian) terms proportional to $c_2$ which are cubic in the gauge fields. In order to match the high-energy couplings in~(\ref{equ:A5}) to the low-energy couplings in~(\ref{equ:A8}), we define
\begin{align}
\Lambda_\gamma^{-1} &= \left( c^2_w \hskip1pt c_a  + s_w^2 \hskip1pt s_a \right) \Lambda^{-1} \, , \label{equ:Lg}\\ 
\Lambda_Z^{-1} &= \left(c^2_w \hskip1pt s_a + s^2_w\hskip1pt c_a\right) \Lambda^{-1} \, , \\
\Lambda_{Z \gamma}^{-1} &= 2 s_w c_w \left(s_a - c_a \right) \Lambda^{-1} \, ,\\
\Lambda_{W}^{-1} &= s_a \hskip 1pt \Lambda^{-1} \, ,
\end{align}
where $\{c_w,s_w\}\equiv \{\cos \theta_w, \sin \theta_w\}$, with $\theta_w \approx 30^\circ$ the Weinberg mixing angle. Using~(\ref{equ:Lg}), we can write~(\ref{equ:Lca}) as a bound on the photon coupling, 
\begin{align}
\Lambda_\gamma(c_a) \,>\,  &\ \left( c^2_w \hskip1pt c_a  + s_w^2  \hskip1pt s_a \right)^{\!-1} \times \left(\frac{\pi^2}{90} g_{*,R}\right)^{\!-1/4} \sqrt{\gamma_{R}(c_a) \hskip1pt \Tr\hskip1pt \Mp} \nonumber\\[2pt]
&\ \equiv\, \lambda_\gamma(\Tr,c_a) \left(\frac{\Tr}{\SI{e10}{GeV}}\right)^{\!1/2} \,.
\label{equ:Lambgamma}
\end{align}
This bound is illustrated in the right panel of Fig.~\ref{fig:ratePhoton}. We see that we get the most conservative constraint by setting $s_a=0$, for which we have $\lambda_\gamma(\SI{e10}{GeV},1) = \SI{1.4e13}{GeV}$.

\subsection{Couplings to Matter Fields}
\label{app:ratesMatter}

The calculation of the Goldstone production rates associated with the couplings to the SM fermions is somewhat less developed.  In this section, we will calculate the relevant rates following the procedure outlined in~\cite{Masso:2002np}.

\paragraph{Preliminaries}	

The integrated Boltzmann equation for the evolution of the number density of the Goldstone boson takes the form
\beq
\dot n_\phi + 3 H n_\phi = \Gamma (n^{\rm eq}_\phi - n_\phi) \, ,
\eeq
where $n^{\rm eq}_\phi = \zeta(3) T^3/\pi^2$ is the equilibrium density of a relativistic scalar. In order to simplify the analysis, we will replace the integration over the phase space of the final states with the center-of-mass cross section, $\sigma_{\rm cm}$, or the center-of-mass decay rate, $\Gamma_{\rm cm}$.  While this approach is not perfectly accurate, it has the advantage of relating the vacuum amplitudes to the thermal production rates in terms of relatively simple integrals.

\begin{itemize}
\item For a two-to-two process, $1+2 \to 3+4$, we have
\beq
\Gamma_{2\to 2} \simeq \frac{1}{n_{\phi}^{\rm eq}} \int \frac{\d^3 p_1}{(2\pi)^3} \frac{\d^3 p_2}{(2\pi)^3} \frac{f_1(p_1)}{2 E_1} \frac{f_2(p_2)}{2 E_2} \big[1\pm f_3\big] \big[1\pm f_4\big]  \,2 s \sigma_{\rm cm}(s)\, ,
\eeq
where $f_{1,2}$ are the distribution functions of the initial states and $s\equiv (p_1+p_2)^2$ is the Mandelstam variable.  We have included simplified Bose enhancement and Pauli blocking terms, $\big[1\pm f_3\big] \big[1\pm f_4\big] \to \frac{1}{2} \big( [1\pm f_3(p_1)] [1\pm f_4(p_2)] +\{ p_1 \leftrightarrow p_2 \} \big)$, which is applicable in the center-of-mass frame where the initial and final momenta are all equal.\footnote{These Pauli blocking and Bose enhancement terms were not included in~\cite{Masso:2002np}, as they complicate the rate calculations.  We have included them to ensure that the rates computed for both the forward and backward processes give the same results.} For $s \gg m^2_i$, the center-of-mass cross section is given by 
\beq
\sigma_{\rm cm}(s) \simeq \frac{1}{32 \pi} \int \d\cos \theta\, \frac{\sum |\M|^2(s, \theta)}{s}  \, ,
\label{equ:SigmaCM}
\eeq
where $\sum |\M|^2$ is the squared scattering amplitude including the sum over spins and charges and $\theta$ is the azimuthal angle in the center-of-mass frame. For all models of freeze-out considered in the main text, the center-of-mass cross section is independent of $s$. In this section, we will only encounter fermion-boson scattering or fermion annihilation. With the enhancement/blocking terms,  one finds that the numerical pre-factors in both cases agree to within 10 percent. To simplify the calculations, we will therefore use the fermion annihilation rate throughout,
\beq
\Gamma_{2\to2} \simeq \sigma_{\rm cm}\,T^3  \left(\frac{7}{8}\right)^2 \frac{\zeta(3) }{ \pi^2}  \approx 0.093 \, \sigma_{\rm cm}\,T^3\, .
\label{eq:Gamma2to2}
\eeq
The advantage of this approach is that we can relate the center-of-mass cross section directly to the production rate with minimal effort and reasonable accuracy.

\item For a one-to-two process, $1 \to 2+3$, the decay rate in the center-of-mass frame is
\beq
 \Gamma_{\rm cm} \simeq \frac{1}{32 \pi \, m_1} \int \d \cos\theta\, \sum |\M|^2\, , \label{equ:A19}
\eeq
where we have taken the two final particles to be massless. Since $\Gamma_{\rm cm}$ is independent of energy, the rate only depends on whether the initial state is a fermion or boson.  Transforming this rate to a general frame gives 
\beq
\Gamma_{1\to 2} \simeq \frac{1}{n_{\phi}^{\rm eq}} \int \frac{\d^3 p_1}{(2\pi)^3} \,f_1(p_1) \big[1\pm f_2(p_1/2)\big]\big[1\pm f_3(p_1/2)\big]\frac{m_1}{E_{1}} \Gamma_{\rm cm}\, , \label{equ:A20}
\eeq
where $f_1$ is the distribution function of the decaying particle (not necessarily $\phi$). We are mostly interested in the limit $T \gg m_1$, in which case the rate (\ref{equ:A20}) reduces to 
\begin{align}
\Gamma_{1\to2} &\simeq  \frac{m_1}{T} \frac{\pi^2}{16 \hskip1pt \zeta(3)} \,\Gamma_{\rm cm} \times \left\{\begin{array}{ll}  \displaystyle  1-\frac{4}{\pi^2} & \qquad {\rm fermion,} \\[14pt] 1 & \qquad {\rm boson,} \end{array} \right.\label{equ:DecayRate}
\end{align}
where the dependence on the number of degrees of freedom of the decaying particle has been absorbed into $\Gamma_{\rm cm}$ through the sum over spins and charges.  Note that, in equilibrium, the rates for decay and inverse decay are equal.
\end{itemize}

\paragraph{Coupling to charged fermions} We consider the following coupling between a Goldstone boson and charged fermions:
\beq
\L_{\phi \psi} =  \frac{\phi}{\Lambda_\psi} \bigg(\i H  \,\bar\psi_{L,i} \!\left[ (\lambda_i - \lambda_j) g_V^{ij} + (\lambda_i + \lambda_j) g_A^{ij} \right]\! \psi_{R,j} + {\rm h.c.} \bigg)\, ,
\label{equ:Lferm}
\eeq
where $H$ is the Higgs doublet, $\psi_{L,R} \equiv \tfrac{1}{2}(1\mp \gamma^5) \psi$, and the $SU(2)_L$ and $SU(3)_c$ structures have been left implicit.  Distinct processes dominate in the various limits of interest:

\begin{itemize}
\item {\bf Freeze-out} \hskip4pt At high energies, the Goldstone boson is produced through the following two processes (see Fig.~\ref{fig:diagramsFermionsHiggs}): ({\it a}\hskip1pt) $\psi_i +\bar \psi_j \to H + \phi$ and ({\it b}\hskip1pt)~$\psi_i +H \to \psi_j +\phi$.
\begin{figure}[t]
\centering
\subfloat[\label{fig:fermionFermionAnnihilationHiggs}Fermion annihilation.]{ 		
\hspace{0.6cm}\includegraphics*{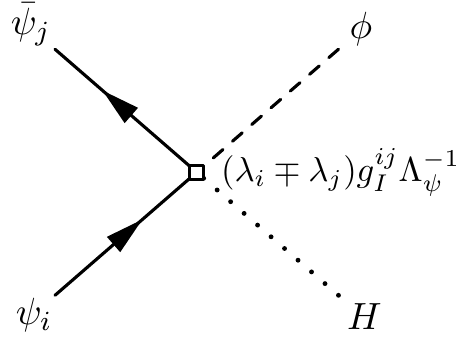}
}\hspace{1.5cm}
\subfloat[\label{fig:fermionFermionHiggs}Fermion-Higgs scattering.]{
\hspace{0.6cm}\includegraphics*{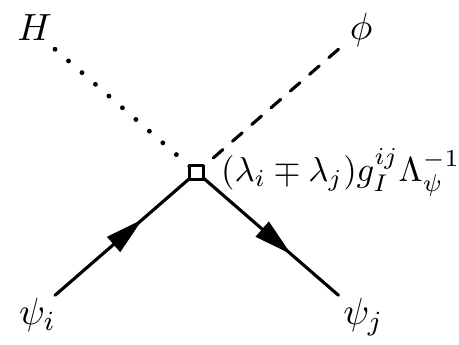}
}
\caption{Feynman diagrams for the dominant Goldstone production via the coupling to charged fermions above the electroweak scale. For the vector and axial vector couplings, $I \in \{V,A\}$, the `$-$' and `$+$' signs apply, respectively. }
\label{fig:diagramsFermionsHiggs}
\end{figure}
Summing over the spins and charges, we get
\begin{align}
\sum |\M|^2_{(a)} &= 4 N_\psi \, s \, \frac{(\lambda_i - \lambda_j)^2 (g_V^{ij})^2 + (\lambda_i + \lambda_j)^2 (g_A^{ij})^2}{\Lambda_\psi^2} \, ,\\
\sum |\M|^2_{(b)} &= 4 N_\psi \, s (1- \cos \theta)  \, \frac{(\lambda_i - \lambda_j)^2 (g_V^{ij})^2 + (\lambda_i + \lambda_j)^2 (g_A^{ij})^2}{\Lambda_\psi^2} \, ,
\end{align}
where we have combined fermion and anti-fermion scattering in the sum over charges and introduced
\beq
N_\psi \equiv  \left\{\begin{array}{ll}  \displaystyle  1 & \qquad \psi = {\rm lepton}, \\[8pt] 3 & \qquad \psi = {\rm quark}. \end{array} \right. 	
\eeq
We also find it convenient to define $\Lambda_{ij}^I \equiv \Lambda_\psi/g^{ij}_I$, with $I \in \{V, A\}$. Using~(\ref{equ:SigmaCM})  and~(\ref{eq:Gamma2to2}), and treating the vector and axial-vector couplings separately, we find
\beq
\Gamma^{I}_{ij} = N_\psi \left(\frac{7}{8}\right)^2 \frac{ 4\hskip1pt \zeta(3)}{\pi^2} \frac{(\lambda_i \mp \lambda_j)^2}{8 \pi } \, \frac{ T^3}{(\Lambda_{ij}^{I})^2} \, \simeq\, 0.19 \hskip1pt N_\psi \frac{(\lambda_i \mp \lambda_j)^2}{8 \pi } \, \frac{ T^3}{(\Lambda_{ij}^{I})^2}\, ,
\eeq
where the `$-$' and `$+$' signs apply to $I=V$ and $I=A$, respectively.  

\item{\bf Freeze-in}  \hskip4pt  Below the scale of EWSB, the Lagrangian~(\ref{equ:Lferm}) becomes
\beq
\L_{\phi \psi} = \i \frac{\phi}{\Lambda_\psi} \bar \psi_i \left[(m_i-m_j) g_V^{ij} +(m_i+m_j) g_A^{ij} \gamma^5 \right] \psi_j\, , \label{equ:Lphipsi}
\eeq
where $m_i \equiv \sqrt{2}\lambda_i/v$.   The Goldstone production processes associated with these couplings are shown in Fig.~\ref{fig:diagramsFermions}.

\vskip4pt
{\it Diagonal couplings.}---We first consider the diagonal part of the interaction, which takes the form $\i \tilde \epsilon_{ii} \,  \phi \hskip1pt \bar \psi_i \gamma^5 \psi_i$, with $\tilde \epsilon_{ii} \equiv 2 m_i g_{A}^{ii} / \Lambda_\psi$.  Kinematical constraints require us to include at least one additional particle in order to get a non-zero amplitude.  The two leading processes are ({\it a}\hskip1pt) $\psi_i  + \{\gamma,g\}\to \psi_i + \phi$ (cf.~Fig.~\ref{fig:fermionComptonProcessS}) and ({\it b}\hskip1pt) $\psi_i + \bar{\psi}_i \to \phi + \{\gamma,g\}$ (cf.~Fig.~\ref{fig:fermionFermionAnnihilation}), where $\{\gamma,g\}$ is either a photon or gluon depending on whether the fermion is a lepton or quark, respectively. 
\begin{figure}[t]
\centering
\subfloat[\label{fig:fermionComptonProcessS}Compton-like process.]{
\includegraphics*{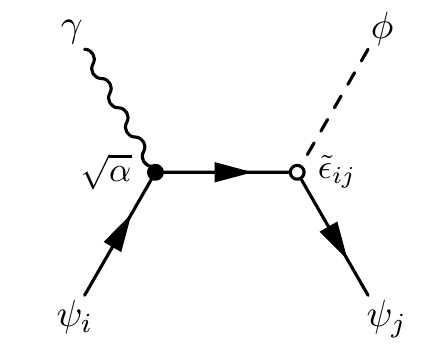}
}\hspace{1.0cm}
\subfloat[\label{fig:fermionFermionAnnihilation}Fermion annihilation.]{
\includegraphics*{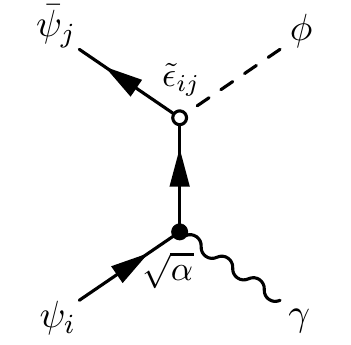}
}\hspace{1.0cm}
\subfloat[\label{fig:fermionFermionDecay}Fermion decay.]{
\includegraphics*{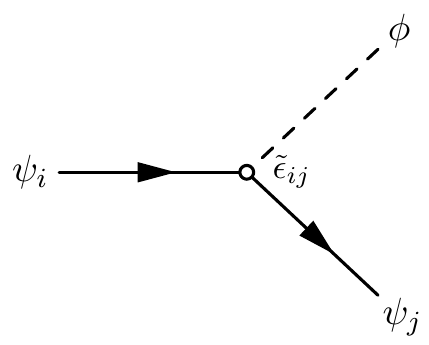}
}
\caption{Feynman diagrams for the dominant Goldstone production via the coupling to charged fermions below the electroweak scale. For quarks, the coupling to photons is replaced by that to gluons. In addition to the displayed $s$- and $t$-channel diagrams for the Compton-like process and fermion annihilation, there are $u$-channel diagrams which are not shown.}
\label{fig:diagramsFermions}
\end{figure} 
Summing over spins and charges, we obtain
\begin{align}
\sum |\M|^2_{(a)} &= 16 \pi \hskip1pt A_\psi  \, |\tilde \epsilon_{ii}|^2  \frac{s^2}{(m_i^2-t)(m_i^2-u)}   \,  ,\\
\sum |\M|^2_{(b)} &= 16 \pi \hskip1pt A_\psi \, |\tilde \epsilon_{ii}|^2 \frac{t^2}{(s-m_i^2)(m_i^2-u)} \, ,
\end{align}
where $s$, $t$ and $u$ are the Mandelstam variables and
\begin{align}
A_\psi &\equiv  \left\{\begin{array}{ll}  \displaystyle  \alpha & \qquad \psi = {\rm lepton}, \\[8pt] 4 \alpha_s & \qquad \psi = {\rm quark}. \end{array} \right. 	
\end{align}
In the massless limit, the cross section has IR divergences in the $t$- and $u$-channels from the exchange of a massless fermion.  The precise production rate therefore depends on the treatment of the soft modes. Regulating the IR divergence with the fermion mass and taking the limit $s \gg m_i^2$, we find
\beq
\sigma_{\rm cm}(s) \simeq \frac{1}{s} \,  A_\psi\, |\tilde \epsilon_{ii}|^2 \left[3 \log \frac{s}{m_i^2} - \frac{3}{2 } \right]  .
\eeq
At high temperatures, the fermion mass is controlled by the thermal mass $m_i^2 \to m_T^2 = \frac{1}{2} \pi A_\psi T^2$ and the production rate becomes
\beq
\tilde \Gamma_{ii} = \frac{3\pi^3}{64\zeta(3)} A_\psi\, \frac{|\tilde \epsilon_{ii}|^2}{8 \pi} T \left[ \log\frac{2}{ \pi A_\psi} +2 \log 2 -\frac{3}{2} \right] .
\label{eqn:chargeddiagonal}
\eeq
This formula is expected to break down at $T \lesssim m_i$, but will be sufficient at the level of approximation being used in this paper.  A proper treatment of freeze-in at $T\sim m_i$ should go beyond $\Gamma =H$ and fully solve the Boltzmann equations.  However, this level of accuracy isn't needed for estimating the constraint on the coupling $\tilde\epsilon_{ii}$.

The result~(\ref{eqn:chargeddiagonal}) will be of limited utility for the coupling to quarks.  This is because, for $T \lesssim \SI{30}{GeV}$, the QCD coupling becomes large and our perturbative calculation becomes unreliable.\footnote{These effects are computable using the techniques of~\cite{Salvio:2013iaa}, but this is beyond the scope of the present work.}  In fact, we see that the production rate~(\ref{eqn:chargeddiagonal}) becomes negative in this regime.    While the top quark is sufficiently heavy to be still at weak coupling, its mass is close to the electroweak phase transition and, therefore, the assumption $s \gg m_t^2$ is not applicable.  For these reasons, we will not derive bounds on the quark couplings from these production rates.

{\it Off-diagonal couplings.}---When the coupling of $\phi$ is off-diagonal in the mass basis, the dominant process at low energies is the decay $\psi_i \to \psi_j +\phi$, cf.~Fig.~\ref{fig:fermionFermionDecay}. Since the mass splittings of the SM fermions are large and $m_\phi \ll m_\psi$, the center-of-mass decay rate is well approximated by
\beq
\Gamma_{\rm cm} = \frac{N_\psi}{8\pi} \frac{m_i^3}{\Lambda_{ij}^2} \, ,
\eeq
where $\Lambda_{ij} \equiv \big[(g_V^{ij})^2 + (g_A^{ij})^2\big]^{-1/2}\, \Lambda_\psi$. Using~(\ref{equ:DecayRate}), we get 
\beq
\tilde \Gamma_{ij} = \frac{(\pi^2-4) }{16\hskip1pt \zeta(3)} \frac{N_\psi}{8\pi} \frac{1}{T} \frac{m_i^4}{\Lambda_{ij}^2} \,\simeq\, 0.31 N_\psi \frac{|\tilde \epsilon_{ij}|^2}{8\pi} \frac{m_i^2}{T} \, ,
\eeq
with $\tilde \epsilon_{ij} \approx m_i/\Lambda_{ij}$. In addition to this decay, we also have production with a photon or gluon, given by~(\ref{eqn:chargeddiagonal}) with $\tilde \epsilon_{ii} \to \tilde \epsilon_{ij}$.   We will neglect this contribution as it is suppressed by a factor of $\alpha$ or $\alpha_s$ for $T \sim m_i$.
\end{itemize}

\paragraph{Coupling to neutrinos}

The coupling between the Goldstone boson and neutrinos is
\beq
\L_{\phi \nu}= - \frac{1}{2}\left( \i \hskip1pt \tilde \epsilon_{ij} \phi \nu_i \nu_j - \frac{1}{2\Lambda_\nu}  \epsilon_{ij}  \phi^2 \nu_i \nu_j  +\cdots \right) + {\rm h.c.} \, , \label{equ:LN}
\eeq
where we have written the Majorana neutrinos in two-component notation.  The first term in~(\ref{equ:LN}) will control freeze-in and the second will determine freeze-out:  

\begin{itemize}
\item{\bf Freeze-out}  \hskip4pt  At high energies, the dominant production mechanism is $\nu_i + \nu_j\to \phi + \phi$ (cf.~Fig.~\ref{fig:neutrinoNeutrinoAnnihilation}) through the second term in $\L_{\nu \phi}$. 
\begin{figure}[t]
\centering
\subfloat[\label{fig:neutrinoNeutrinoAnnihilation}Neutrino annihilation.]{
\includegraphics*{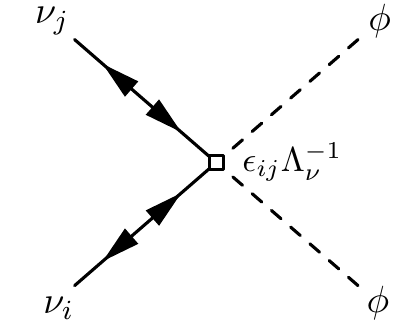}
}\hspace{1.0cm}
\subfloat[\label{fig:neutrinoNeutrinoDecay}Neutrino decay.]{
\includegraphics*{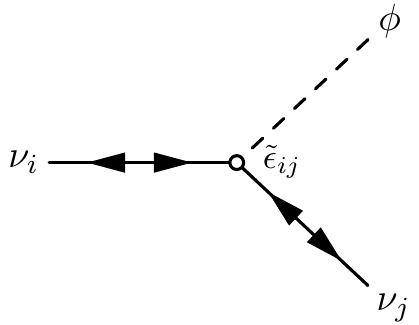}
}\hspace{1.0cm}
\subfloat[\label{fig:neutrinoInverseDecay}Inverse decay.]{
\includegraphics*{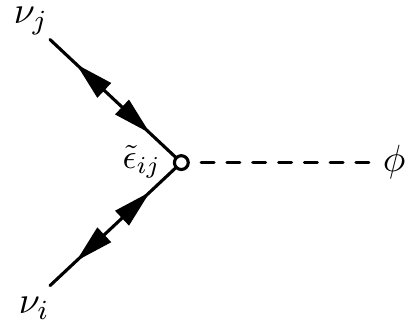}
}
\caption{Feynman diagrams for the dominant Goldstone production via the coupling to neutrinos.  The double arrows denote the sum over spinor index structure for two-component fermions~\cite{Dreiner:2008tw}.}
\label{fig:diagramsNeutrinos}
\end{figure} 
The spin-summed amplitude squared is
\beq
\sum |\M|^2 = |\epsilon_{ij}|^2 \frac{2 s}{\Lambda_\nu^2}\, ,
\eeq  
which results in the production rate
\beq
\Gamma_{ij} = \frac{1}{2}s_{ij}\left(\frac{7}{8}\right)^2 \frac{\zeta(3)}{\pi^2}\, \frac{|\epsilon_{ij}|^2}{8 \pi} \frac{T^3}{\Lambda_\nu^2}  \,\simeq\, 0.047 \hskip1pt s_{ij}\, \frac{|\epsilon_{ij}|^2}{8 \pi} \frac{T^3}{\Lambda_\nu^2}\, ,
\eeq
where the factors of $\frac{1}{2}$ and $s_{ij} \equiv 1-\frac{1}{2} \delta_{ij}$ are the symmetry factors for identical particles in the initial and final states, respectively. The contribution to the rate from higher-order terms in~(\ref{equ:LN}) is suppressed by further powers of $T^2/\Lambda_\nu^2$.

\item{\bf Freeze-in} \hskip4pt Unlike for charged fermions, the freeze-in abundance from the coupling to neutrinos arises only through decays.  Below the scale of EWSB, the  couplings of neutrinos to the rest of the SM are suppressed by the weak scale and are irrelevant.  The only freeze-in processes that are allowed by kinematics are therefore three-body decays.  

\vskip4pt
{\it Low-mass regime.}---For $m_\phi \ll m_i-m_j$, with $m_i > m_j$, the off-diagonal linear coupling allows the decay $\nu_i \to \nu_j +\phi$, cf.~Fig.~\ref{fig:neutrinoNeutrinoDecay}. The decay rate in the center-of-mass frame is
\beq
\Gamma_{\rm cm} = \frac{1}{8\pi} \frac{m_i^2-m_j^2}{m_i^3} \left(|\tilde \epsilon_{ij}|^2  (m_i^2+m_j^2) + 2\,\mathrm{Re}\!\left[\!(\tilde \epsilon_{ij})^2\right] m_i m_j  \right) .
\eeq
In order to simplify the calculations in the main text, we take $m_i \gg m_j$ which is guaranteed for the minimal mass normal hierarchy.  Since the decaying particle is a fermion, the thermal production rate in~(\ref{equ:DecayRate}) becomes
\beq
\tilde \Gamma_{ij} = \frac{\pi^2-4}{16\hskip1pt\zeta(3)}  \frac{|\tilde \epsilon_{ij}|^2 }{8\pi} \frac{m_{i}^2}{T} \,\simeq\, 0.31 \frac{|\tilde \epsilon_{ij}|^2 }{8\pi} \frac{m_{i}^2}{T} \, .
\eeq
Notice that the off-diagonal decay rate is the same for charged leptons and neutrinos even though the neutrinos have a Majorana mass.

\vskip4pt
{\it High-mass regime.}---For $m_\phi \gg m_i \ge m_j$, the Goldstone boson decays to fermions, $\phi \to \nu_i + \nu_j$, both through the diagonal and off-diagonal couplings. The inverse decay $\nu_i + \nu_j \to \phi$ (see Fig.~\ref{fig:neutrinoInverseDecay}) is therefore a production channel. The decay rate is given by
\beq
\Gamma_{\rm cm} = \frac{|\tilde \epsilon_{ij}|^2 }{8\pi} m_{\phi} \, ,
\eeq
which, in equilibrium, is equal to the rate for the inverse decay. Since the decaying particle is a boson, the thermal production rate in~(\ref{equ:DecayRate}) becomes
\beq
\tilde \Gamma_{ij} =  s_{ij} \, \frac{\pi^2}{16\hskip1pt \zeta(3)}  \frac{|\tilde \epsilon_{ij}|^2 }{8\pi} \frac{m_{\phi}^2}{T}  \,\simeq\, 0.51 \hskip1pt s_{ij}\, \frac{|\tilde \epsilon_{ij}|^2 }{8\pi} \frac{m_{\phi}^2}{T} \, .
\eeq
The rate is somewhat enhanced compared to the decay in the low-mass regime because the decaying particle is a boson.

\end{itemize}
The dominant Goldstone production mechanism through the couplings to neutrinos is quite sensitive to kinematics.  For $m_\phi \lesssim m_{\nu}$, the diagonal decay is forbidden and the dominant Goldstone production is through freeze-out.  In addition, when $m_\phi \sim m_\nu$ there are additional kinematical constraints for both diagonal and off-diagonal couplings.  As a result, the limits on the interaction scale $\Lambda_\nu$ (or the dimensionless couplings $\epsilon_{ij}$ and $\tilde \epsilon_{ij}$) will be sensitive to $m_\phi$.

\clearpage
%%%%%%%%%%%%%%%%
\section{Comments on Decays}
\label{app:decays}
%%%%%%%%%%%%%%%%

Throughout the paper, we have treated each of the operators which couple the pNGBs to the SM independently. For computing the production rates, this is justified since the amplitudes for the different processes that we consider do not interfere and the couplings therefore add in quadrature.  One may still ask, however, if the interplay between several operators can affect the cosmological evolution after the production.  In particular, one might worry that some operators would allow for the decay of the pNGBs and that this might evade the limits on $\Neff$.  In this appendix, we will address this concern. We are assuming that $m_\phi < \SI{1}{MeV}$, so that the only kinematically allowed decays are to photons and neutrinos.

\paragraph{Decay to photons}
If the decay occurs after recombination, then the pNGBs are effectively stable as far as the CMB is concerned and our treatment in the main text applies directly.  To see when this is the case, we computed the decay temperature $\Td$ associated with the decay mediated by the coupling to photons~(\ref{equ:phiF}):
\beq
\frac{\Td}{\Trec} \,\approx\, \num{9.5e-10} \left(\frac{\Lambda_\gamma}{\SI{e10}{GeV}} \right)^{\!-4/3} \left(\frac{m_\phi}{\Trec}\right)^{\!2} \, .
\eeq
Recalling the stellar cooling bound, $\Lambda_\gamma >\SI{1.3e10}{GeV}$~\cite{Friedland:2012hj}, we see that the pNGBs are effectively stable as long as $m_\phi\lesssim \SI{10}{keV}$.  For comparison, a stable particle with $m_\phi \gtrsim \SI{100}{eV} $ produces $\Omega_m >1$ and is therefore excluded by constraints on the dark matter abundance.  For $m_\phi>\SI{10}{keV}$, the decay to photons does affect the phenomenology and must be considered explicitly.  Nevertheless, in the regime of interest, the pNGBs are non-relativistic and, therefore, carry a large energy density, $\rho_\phi \simeq m_\phi n_\phi$.  As a result, this region is highly constrained by current cosmological observations~\cite{Cadamuro:2011fd, Millea:2015qra}.

\paragraph{Decay to neutrinos}
Depending on the mass of the pNGB, the decay to neutrinos leads to the following three scenarios: 
\begin{itemize}
\item For $m_\phi < \Trec$, the implications of the decays are relatively easy to characterize.  As discussed in \S\ref{sec:nufreezein}, the phenomenology is only modified if $\Tfl>\Trec$. In this case, strong interactions between the pNGBs and the neutrinos imply that the neutrinos are no longer free-streaming particles, which is ruled out by recent CMB observations~\cite{Baumann:2015rya}.

\item For $\Td> m_\phi > \Trec$, the pNGBs are brought into equilibrium with the neutrinos at $T \sim \Td$ and then become Boltzmann suppressed for $T\lesssim m_\phi$.  This process leads to a contribution to $\Neff$, even if the pNGBs have negligible energy density to begin with.  To estimate the size of the effect, we first note that the freeze-in at $\Td$ conserves the total energy density in neutrinos and pNGBs, 
\beq
(g_{\nu} + g_\phi) (a_1 T_1)^4 = g_{\nu}  (a_0 T_0)^4\, ,
\eeq
where $T_0$ and $T_1$ are the initial and final temperatures during the equilibration, and $g_{\nu}$ and $g_\phi =1$ are the effective numbers of degrees of freedom in $\nu$ and $\phi$, respectively.  When the temperature drops below the mass of the pNGBs, their energy density is converted to neutrinos.  This process conserves the comoving entropy density,
\beq
(g_{\nu} + g_\phi) (a_1 T_1)^3 =g_{\nu}   (a_2 T_2 )^3 \, ,
\eeq
where $T_2 \ll m_\phi$ is some temperature after the pNGB population has decayed. The final energy density of the neutrinos becomes
\beq
a_2^4 \hskip1pt \rho_{\nu,2} = \left(\frac{g_{\nu} + g_\phi}{g_{\nu}} \right)^{\!1/3} a_0^4\hskip1pt \rho_{\nu, 0} \, ,
\eeq
where $\rho_{\nu, i}\equiv \rho_\nu(a_i)$.  Using the definition of $\Neff$ in~(\ref{equ:Neff}) and $a^4 \rho_\gamma=const.$, we find 
\beq
\Neff =\left(\frac{g_{\nu} + g_\phi}{g_{\nu}} \right)^{\!1/3} \Neff{}_{,0}\, .
\eeq
Considering the coupling to a single neutrino flavor (rather than all three), i.e.~$\Neff{}_{,0} \simeq 1$ and $g_\nu = 7/4$, we then get
\beq
\Delta\Neff = \left(1+ \frac{4}{7}\right)^{\!1/3} -1 \,\simeq\, 0.16\, ,
\eeq
where $\Delta \Neff \equiv \Neff-\Neff{}_{,0}$.  Coupling to more than one neutrino flavor and including a non-zero initial temperature for the pNGBs would increase this number slightly, so that we will use $\Delta \Neff \geq 0.16$.

\item The production of pNGBs through the freeze-in process is avoided if $m_\phi > \Td > \Trec$, in which case the pNGBs decay to neutrinos out of equilibrium.  To a good approximation, this decay conserves the energy density, which is therefore simply transferred from $\phi$ to $\nu$ at the time of the decay.  The contribution to $\Delta \Neff$ is enhanced by the amount of time that $\phi$ is non-relativistic before its decay, which may be a large effect for $m_\phi \gg 1$ eV (see e.g.~\cite{Fischler:2010xz} for a related discussion).
\end{itemize}
In summary, operators that allow the Goldstone bosons to decay do not substantially alter the predictions presented in the main text.  On the one hand, decays to photons cannot occur early enough to impact the CMB.  On the other hand, decays to neutrinos typically increase the contributions to $\Delta \Neff$ and would therefore strengthen our bounds.

\clearpage
\phantomsection
\addcontentsline{toc}{section}{References}
\bibliographystyle{utphys}
\bibliography{Refs}
\end{document}